\documentclass[showpacs,preprintnumbers,amsmath,amssymb,floatfix]{revtex4}
\usepackage{xcolor}
\usepackage{graphicx}
\usepackage{epsfig}
\usepackage{amsfonts}
\usepackage{amssymb}
\usepackage{epsf}
\newcommand{\insertplot}[5]{\begin{figure}
 \hfill\hbox to 0.05in{\vbox to #5in{\vfill
 \inputplot{#1}{#4}{#5}}\hfill}
 \hfill\vspace{-.1in}
 \caption{#2}\label{#3}
 \end{figure}}
\newcommand{\inputplot}[3]{
 \special{ps: plotfile #1}
}

\newcounter{fig}

\begin{document}

\title{
Innermost stable circular orbits of neutron stars 
in dilatonic-Einstein-Gauss-Bonnet theory}

\author{
Guillermo A. Gonz\'alez$^1$, Burkhard Kleihaus$^2$,  
Jutta Kunz$^2$ and Sindy Mojica$^1$}

\affiliation{
$^1${Grupo de Investigaci\'on en Relatividad y Gravitaci\'on,  
Escuela de F\'isica,}\\
  {\small  Universidad Industrial de Santander, 
A. A. 678, Bucaramanga 680002, Colombia.}\\
$^2${Institut f\"ur Physik, Universit\"at Oldenburg,
D-26111 Oldenburg, Germany}
}
\date{\today}

\begin{abstract}
The innermost stable circular orbits (ISCOs) around 
rapidly rotating neutron stars are studied
in dilatonic Einstein-Gauss-Bonnet theory. 
Universal relations for properly scaled ISCO properties
are extended from General Relativity to 
dilatonic Einstein-Gauss-Bonnet theory
and additional relations are obtained.
\end{abstract}
\maketitle


\section{Introduction}

Currently General Relativity (GR) 
fully describes the behaviour of the gravitational force, 
as tested by experiments and observations in the solar system and beyond,
involving both weak and strong gravitational fields
\cite{Capozziello:2011et,Will:2014kxa,Berti:2015itd}.
However, in the quest of a better understanding of all fundamental forces 
of nature, gravity is still the only one without an appropriate description 
at the quantum level. 
Thus, although GR is continuing to pass all tests, there are various attempts 
to formulate a quantum theory of gravity.

Here one of the popular contenders is String Theory,
which in its low energy limit has led to 
effective field theories, whose predictions can be tested
analogously to those of GR
\cite{Capozziello:2011et,Will:2014kxa,Berti:2015itd}.
String Theory predictions include the presence of higher curvature terms
and additional fields \cite{Zwiebach:1985uq,Gross:1986mw,Metsaev:1987zx}.
Among the String Theory motivated alternative theories of gravity 
dilatonic Einstein-Gauss-Bonnet theory (dEGB) has led to particular interest 
because it yields only second order equations of motion,
and thus it is ghost free. 

Astrophysical compact objects are an important probe to test the range
of validity of GR and alternative theories of gravity like dEGB, 
because their strong gravitational fields may provide information 
on how gravity behaves under extreme conditions. Therefore deviations
from the predictions of GR may arise due to the presence of
higher curvature terms and additional fields. In dEGB theory
the higher curvature terms enter via the Gauss-Bonnet (GB) term,
that is coupled to a scalar dilaton field,
see $e.g.$ \cite{Moura:2006pz}.
EGBd theory may also be considered as a
particular case of Horndeski gravity \cite{Kobayashi:2011nu}.

Black hole solutions have been studied in dEGB theory for quite some time
\cite{Mignemi:1992nt,Mignemi:1993ce,Kanti:1995vq,Torii:1996yi,
Alexeev:1996vs,Guo:2008hf,Pani:2009wy,Pani:2011gy,Kleihaus:2011tg,
Ayzenberg:2014aka,Maselli:2015tta,Kleihaus:2015aje}
starting with the perturbative solutions, and then extending them to fully 
nonperturbative static and rotating black hole solutions.
Interesting features of these dEGB black holes include a minimal mass for
fixed coupling constants, and a violation of the Kerr limit,
i.e., angular momenta exceeding $J/M^2=1$.

Neutron stars, on the other hand, are not as simple objects as black holes, 
since the interior solutions of the Einstein field equations 
must be taken into account, as well. This means that the 
stress-energy tensor contributes decisively, when obtaining the solutions 
of the Einstein equations, and an equation of state (EOS) describing 
the interior structure must be included. 
However, the EOS of matter under such extreme conditions
as found inside a neutron star is yet unknown \cite{Lattimer:2012nd}.

This current uncertainty in the EOS is mainly due to the limitations 
of Earth based laboratories to achieve the huge densities 
present in neutron star cores. 
Indeed, there exists a large body of literature on the neutron star
EOS \cite{Lattimer:2012nd,Ozel:2016oaf,Baym:2017whm}. 
These EOS range from purely hadronic matter to hybrid quark-hadron
matter, all the way to pure quark matter, and thus giving rise to
quark stars.
The choice of the EOS employed then dictates the physical properties
of these compact stars, such as their radii and masses.
Observations of neutron star masses on the order of $M \approx 2 M_\odot$
\cite{Demorest:2010bx,Antoniadis:2013pzd}
thus provide a strong selection criterion for viable EOS.

A valuable tool to further our understanding of the properties of
neutrons stars are their so-called universal relations.
Such universal relations represent intriguing relations between 
appropriately scaled physical properties of neutron stars, which 
are almost independent of the EOS
(see e.g.~the recent reviews \cite{Yagi:2016bkt,Doneva:2017jop}).
In these universal relations compactness plays an important role.
Most prominent among these relations are the $I$-Love-$Q$ relations
of neutron stars between their moments of inertia $I$, their Love numbers,
and their quadrupole moments $Q$
\cite{Yagi:2013bca,Yagi:2013awa,Doneva:2013rha,Pappas:2013naa,
Chakrabarti:2013tca,Yagi:2014bxa,Yagi:2014qua}.

While neutron stars and their properties have been widely studied
in GR, also in the rapidly rotating case
\cite{Hartle:1967he,Cook:1993qj,Cook:1993qr,Stergioulas:2003yp,
Gourgoulhon:2010ju,Friedman:2013xza,Cipolletta:2015nga},
much less is known about neutron stars in alternative
theories of gravity. A recent review on the state of the art 
can be found in \cite{Berti:2015itd}. 
In dEGB theory neutron stars have been investigated in
\cite{Pani:2011xm,Pani:2014jra,Kleihaus:2014lba,Kleihaus:2016dui},
including both the static and the rapidly rotating case,
with associated $I$-$Q$ relations,
yielding little deviation from the well-known GR case
(similarly to the case of scalar-tensor theory \cite{Doneva:2014faa}).

In this paper, rotating neutron stars in dEGB are considered for two EOS
referred to as DI-II \cite{Diaz-Alonso:1985} 
and FPS \cite{Lorenz:1992zz,Haensel:2004nu}, 
for which the $2M_\odot$ limit 
\cite{Demorest:2010bx,Antoniadis:2013pzd}
is only reached and exceeded for fast rotating neutron stars
\cite{Kleihaus:2014lba,Kleihaus:2016dui}.
Various properties of these dEGB neutron stars have been discussed before
\cite{Pani:2011xm,Pani:2014jra,Kleihaus:2014lba,Kleihaus:2016dui}.
In particular, it has been shown, that with increasing GB coupling constant
the maximum mass of the neutron stars decreases.

However, an analysis for test particles surrounding 
a rotating dEBG neutron star has not been performed so far,
although such an analysis has been made for black holes in dEGB 
(see \cite{Kleihaus:2011tg,Kleihaus:2015aje}).
It is therefore a natural important next step to investigate 
the innermost stable circular orbit (ISCO) 
\cite{Carter:1968rr,Bardeen:1970}
for dEGB neutron stars following a similar procedure. 
Despite the numerical limitations, this may in fact reveal
significant effects of the scalar field 
on test particles orbiting around a rotating dEBG neutron star.

In GR the ISCO has been studied for numerous
rotating neutron star models and EOSs 
(see e.g., \cite{Miller:1998gr,Shibata:1998xw,Zdunik:2000qn,
GondekRosinska:2000mp,Pachon:2006an,Bhattacharyya:2011wm,
Pappas:2012nv,Pappas:2012nt, Gondek-Rosinska:2014aaa,
Torok:2014ina,Cipolletta:2016yqv,Luk:2018xmt}).
Concerning alternative theories of gravity the ISCO for
rotating neutron stars has been addressed
in \cite{Pani:2011xm,Doneva:2014uma,Staykov:2015kwa}.
Important astrophysical applications of the ISCO concern
the description and understanding of
accretion disks around neutron stars.
Moreover,
it has been suggested that the ISCO has a direct relation 
with quasi-periodic oscillations (QPOs). 
This phenomenon occurs in x-ray binaries 
and is considered a relevant neutron star signature. 
In fact, the ISCO may be interpreted as the upper bound 
of the observed frequency provided by x-ray observations 
\cite{vanderKlis:2005,vanDoesburgh:2018oom}.

In section II we present the theoretical setting for
our investigations, including the dEGB action, the metric,
the EOS, and the equations of motion for the ISCOs.
We discuss our results in section III, starting
with a brief recollection of the domain of existence of
the dEGB neutron stars. We then present the ISCOs
for these neutron star models, and address their universal relations.
We present our conclusions in section IV.

\section{Theoretical setting}

\subsection{dEGB gravity}

The action describing dilatonic-Einstein-Gauss-Bonnet gravity is given by
\begin{eqnarray}
S=\frac{1}{16 \pi G}\int d^4x \sqrt{-g} 
\left[R -\frac{1}{2}(\partial_\mu \phi)^2  
+ \alpha e^{-\gamma\phi} R^2_{\rm GB}\right] 
+ S_{\rm matter} ,
\label{act}
\end{eqnarray}
where $G$ is Newton's constant, $R$ is the curvature scalar, 
$\phi$ denotes the dilaton field, 
$\alpha$ and $\gamma$ are coupling constants, 
$R_{GB}^2$ denotes the Gauss-Bonnet term, 
$$\label{RGB}
R_{GB}^2=R_{\mu \nu \rho \sigma}R^{\mu \nu \rho \sigma}
-4R_{\mu \nu}R^{\mu \nu}+R^2 ,
$$
while the action for the nuclear matter
is symbolized by $S_{\rm matter}$.

The gravitational field equations are then given by
\begin{eqnarray}
G_{\mu\nu} & = &
\frac{1}{2}\left[\nabla_\mu \phi \nabla_\nu \phi 
                 -\frac{1}{2}g_{\mu\nu}\nabla_\lambda \phi \nabla^\lambda\phi 
		 \right]
\nonumber\\
& &
-\alpha e^{-\gamma \phi} 
\left[	H_{\mu\nu}
  +4\left(\gamma^2\nabla^\rho \phi \nabla^\sigma \phi
           -\gamma \nabla^\rho\nabla^\sigma \phi\right)	P_{\mu\rho\nu\sigma}
		 \right]
\nonumber\\
& &
		 +8\pi G T_{\mu\nu},
\label{eom1}
\end{eqnarray}
where
\begin{eqnarray}
H_{\mu\nu} & = & 2\left[R R_{\mu\nu} -2 R_{\mu\rho}R^\rho_\nu
                        -2 R_{\mu\rho\nu\sigma}R^{\rho\sigma}
			+R_{\mu\rho\sigma\lambda}R_\nu^{\ \rho\sigma\lambda}
		   \right]
		   -\frac{1}{2}g_{\mu\nu}R^2_{\rm GB},
\label{eom2}
\end{eqnarray}
\begin{eqnarray}
P_{\mu\nu\rho\sigma} & = & 
R_{\mu\nu\rho\sigma}
+2 g_{\mu [ \sigma} R_{\rho ]\nu}
+2 g_{\nu [ \rho} R_{\sigma ]\mu}
+R g_{\mu [ \rho} g_{\sigma ]\nu}.
\label{eom3}
\end{eqnarray}
Note that in four dimensions $H_{\mu \nu}=0$. 

The stress-energy tensor $T_{\mu\nu}$  of the nuclear matter
occurring on the right hand side of the field equations is
taken to be in the form of a perfect fluid
\begin{equation}
T_{\mu\nu} = \left(\varepsilon + P\right) U_\mu U_\nu + P g_{\mu\nu} 
\label{tmunu}
\end{equation}
with energy density $\varepsilon$, pressure $P$,
and four velocity $U_\mu$ of the fluid.
In hydrostatic equilibrium the stress-energy tensor 
is covariantly conserved,
\begin{equation}
\nabla_\mu T^{\mu\nu} = 0 \ .
\label{Tpde}
\end{equation}

Finally, the dilaton field equation is given by
\begin{eqnarray}
\nabla^2\phi = \gamma\alpha  e^{-\gamma\phi} R^2_{\rm GB}.
\label{eomd4}
\end{eqnarray}
For the GB coupling constant $\alpha$ 
three values are chosen,
$\alpha=0$ (GR limit), $\alpha=1$ and $\alpha=2$,
while for the dilaton coupling constant $\gamma$
the string value $\gamma=1$ is fixed
(see our previous study for details \cite{Kleihaus:2016dui},
where also observational restrictions for the coupling parameter
$\alpha$ are discussed, following \cite{Pani:2011xm,Yagi:2012gp}).

\subsection{Metric and fluid}

The metric describing the stationary, axially symmetric spacetime 
is specified in terms of the spherical coordinates $r$ and $\theta$,
and reads in quasi-isotropic metric \cite{Kleihaus:2000kg}
\begin{equation}
ds^2 = g_{\mu\nu}dx^\mu dx^\nu
= - e^{2 \nu_0}dt^2
      +e^{2(\nu_1-\nu_0)}\left(e^{2 \nu_2}\left[dr^2+r^2 d\theta^2\right] 
       +r^2 \sin^2\theta
          \left(d\varphi-\omega  dt\right)^2\right) 
\label{metric} 
\end{equation}
with metric functions $\nu_0(r,\theta)$, $\nu_1(r,\theta)$, 
$\nu_2(r,\theta)$ and $\omega(r,\theta)$. 

For a uniformly rotating neutron star fluid
the four velocity has the form
\begin{equation}
U^\mu = \left( u, 0, 0, \Omega u\right) \ , 
\label{Umu}
\end{equation}
where the constant angular velocity of the fluid
is denoted by $\Omega$.
The normalization condition for the four
velocity of the fluid $U^\mu U_\mu = -1$ 
can then be utilized to express the velocity function
$u$ in terms of the metric functions
$\nu_0$, $\nu_1$ and $\omega$ and the fluid
angular velocity $\Omega$,
\begin{equation}
u^2 = \frac{e^{-2\nu_0}}{1-(\Omega-\omega)^2 r^2\sin^2\theta e^{2\nu_1-4\nu_0}}  .
\label{u2}
\end{equation}

The set of equations (\ref{Tpde}), $\nabla_\mu T^{\mu\nu}= 0$,
result in the differential equations for the pressure $P$
\begin{equation}
\frac{\partial_r P}{\varepsilon +P}  =  \frac{\partial_r u}{u} 
 , \ \ \
\frac{\partial_\theta P}{\varepsilon +P}  =  \frac{\partial_\theta u}{u} ,
\label{dpres}
\end{equation}
and need to be supplemented by an EOS,
$\varepsilon = \varepsilon(P)$,
relating the energy density $\epsilon$ to the pressure.

In a polytropic EOS the pressure $P$ is
related directly to the baryon mass density $\rho$ via
\cite{Friedman:2013xza}
\begin{equation}
P=K \rho^{\Gamma} \ , \ \ \ \Gamma = 1 + \frac{1}{N}
\label{polEOSprho}
\end{equation}
where $K$ is the polytropic constant $K$, 
$\Gamma$ the polytropic exponent $\Gamma$,
and $N$ the polytropic index $N$.
The energy density $\varepsilon$ of the fluid is then given by
\begin{equation}
\varepsilon = NP + \rho .
\end{equation}
Based on the calculations presented in \cite{Kleihaus:2016dui},
we here have taken two EOS, 
the polytropic DI-II EOS \cite{Diaz-Alonso:1985} 
with polytropic index $N=0.7463$ and polytropic number $K=1189.0$, 
and the FPS EOS \cite{Lorenz:1992zz,Haensel:2004nu} 
with polytropic parameters $N=0.6104$ and $K=5392.0$.

\subsection{Innermost stable circular orbits}

The stability of circular orbits is physically relevant 
for studies of accretion disks of particles 
orbiting around compact objects such as black holes or neutron stars. 
Instabilities due to perturbations of the circular orbits 
in the equatorial plane and perpendicular perturbations are decoupled. 
In this paper our interest is focused 
only on the first case \cite{Bardeen:1970}. 

The general set of geodesics is obtained from the Lagrangian
\begin{equation}
2 {\mathcal L} = e^{-2 \beta \phi} g_{\mu\nu}\dot x^\mu  \dot x^\nu ,
\label{lag2}
\end{equation}
together with the normalization condition 
\begin{equation}
e^{-2 \beta \phi} g_{\mu\nu}\dot x^\mu  \dot x^\nu =-\epsilon,
\label{norm4v}
\end{equation}
where $\epsilon=0$ and 1 for massless and massive particles, respectively,
and the coupling between matter and the dilaton field is associated
with the coupling constant $\beta$, which in heterotic string theory 
assumes the value $\beta=0.5$ \cite{Pani:2009wy}.
The derivative with respect
to the affine parameter along the geodesics is denoted by a dot.

Explicitly we find for massive particles orbiting in the equatorial plane
($\theta = \pi/2$) the Lagrangian $\mathcal{L}$, resp.  normalization condition
\begin{eqnarray}
2 \mathcal{L} & = & e^{-2 \beta \phi} \left[
(-e^{2\nu_0} + \omega^2 r^2 e^{2(\nu_1-\nu_0)})\dot{t}^2 
+ e^{2(\nu_1 -\nu_0 +\nu_2)} \dot{r}^2
 +e^{2(\nu_1 -\nu_0)}r^2  \dot{\varphi}^2 
-2\omega e^{2(\nu_1 -\nu_0)}r^2\dot{t}\dot{\varphi} \right] 
\\
-1 & = & e^{-2 \beta \phi} g_{\mu\nu}\dot x^\mu  \dot x^\nu \ .
\label{norm4vx}
\end{eqnarray}
Stationarity and axial symmetry imply the existence of two Killing vectors 
associated with two conserved quantities, the particle energy $E$ 
and angular momentum $L$, 
\begin{eqnarray}
E & = & -\frac{\partial \mathcal{L} }{\partial \dot{t}} \\
L & = &  \frac{\partial \mathcal{L} }{\partial \dot{\varphi}}
\end{eqnarray}

The full set of equations of motion for a massive particle orbiting 
in the equatorial plane then becomes
\begin{eqnarray}
\dot{t} & = & \left(E-\omega L\right) e^{-2\nu_0} e^{2\beta \phi}
\end{eqnarray}
\begin{eqnarray}
\dot{\varphi} & = & 
\left(
\omega (E -\omega L)e^{-2\nu_0}  + \frac{L}{r^2}e^{2( \nu_0- \nu_1)}
\right) e^{2\beta \phi}
\end{eqnarray}
\begin{eqnarray}\label{pot}
\dot{r}^2 =
\frac{e^{4\beta \phi}}{e^{2( \nu_1+\nu_2)}} 
\left( E - V_+\right)\left( E - V_-\right)  \equiv V(r)\ , \ {\rm with} \
V_\pm  =  
\omega L + \sqrt{e^{2 \nu_1}e^{-2\beta \phi}+\frac{L^2}{r^2}e^{2 \nu_0}} \ 
 e^{\nu_0-\nu_1} \ ,            
\end{eqnarray}
where for the last equation the normalization condition Eq.~(\ref{norm4vx})
has been used.

The orbital angular velocity of the particle is obtained via
 \begin{equation}
\label{ISCO12}
\Omega_c=\frac{\dot \varphi}{\dot t}
=\omega+\frac{1}{r^2}\frac{L e^{4\nu_0-2\nu_1}}{E-L\omega}.
\end{equation}
We note that for circular orbits the potential satisfies $V(r)=V'(r)=0$.
These conditions yield two relations for the particle energy $E$
and angular momentum $L$ to be solved.
The resulting two sets of solutions 
correspond to corotating and counterrotating orbits.

In order to obtain stable orbits, the second derivative
of the effective potential should be required to be negative,
$V''(r) < 0$. 
At a particular value of the radial coordinate $r=r_{\rm ISCO}$
the stability changes,
with the stable circular orbits residing at $r>r_{\rm ISCO}$.

\section{Results}

In the following we first recall the properties of the dEGB and GR
neutron star models. Then we discuss the ISCO radii and frequencies
for these models. Subsequently we address universal relations
for the ISCOs in GR and dEGB theory.

In the following we will adapt units such that $\Omega$ and $\Omega_c$
are in units of $c/R_0 = 2.031\times 10^5 s$, unless otherwise stated.

\subsection{Neutron star models and ISCO ranges}

\begin{figure}[h!]
\begin{center}

(1.1)\includegraphics[height=.23\textheight, angle =0]{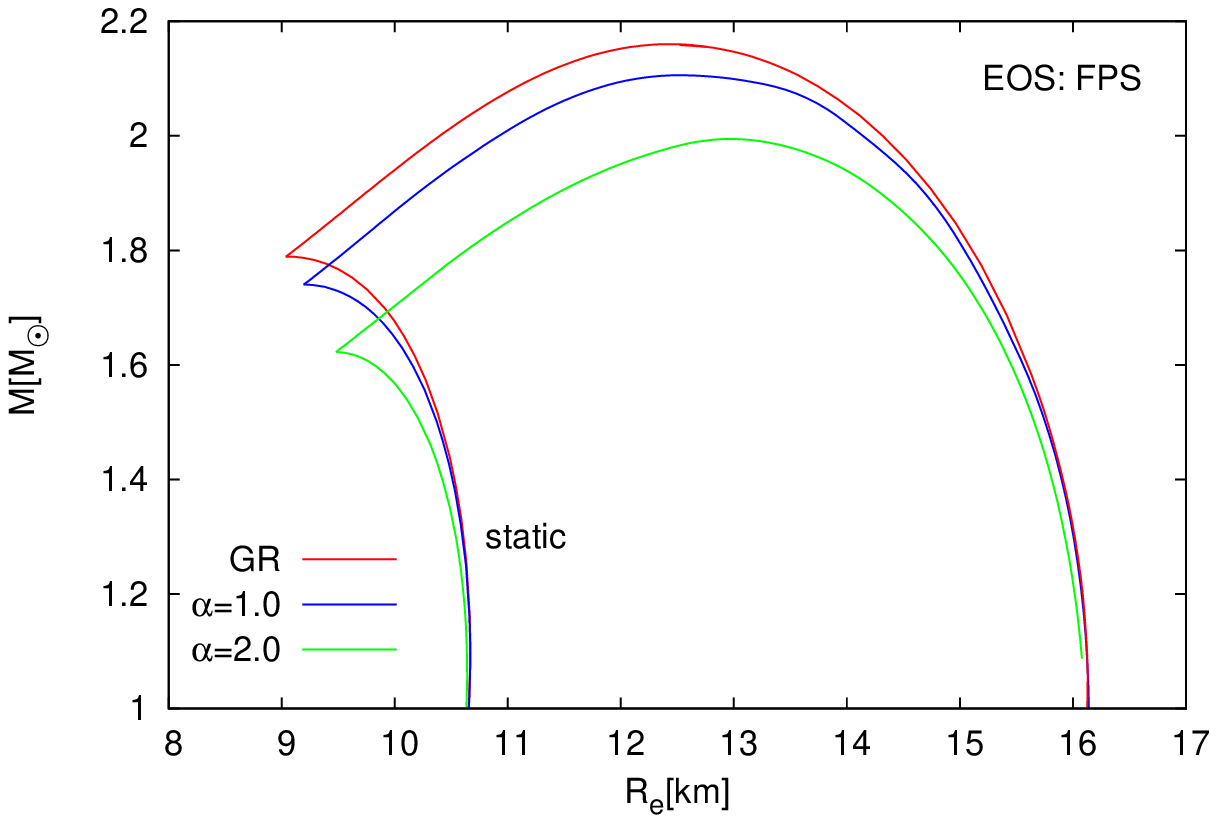}
(1.2)\includegraphics[height=.23\textheight, angle =0]{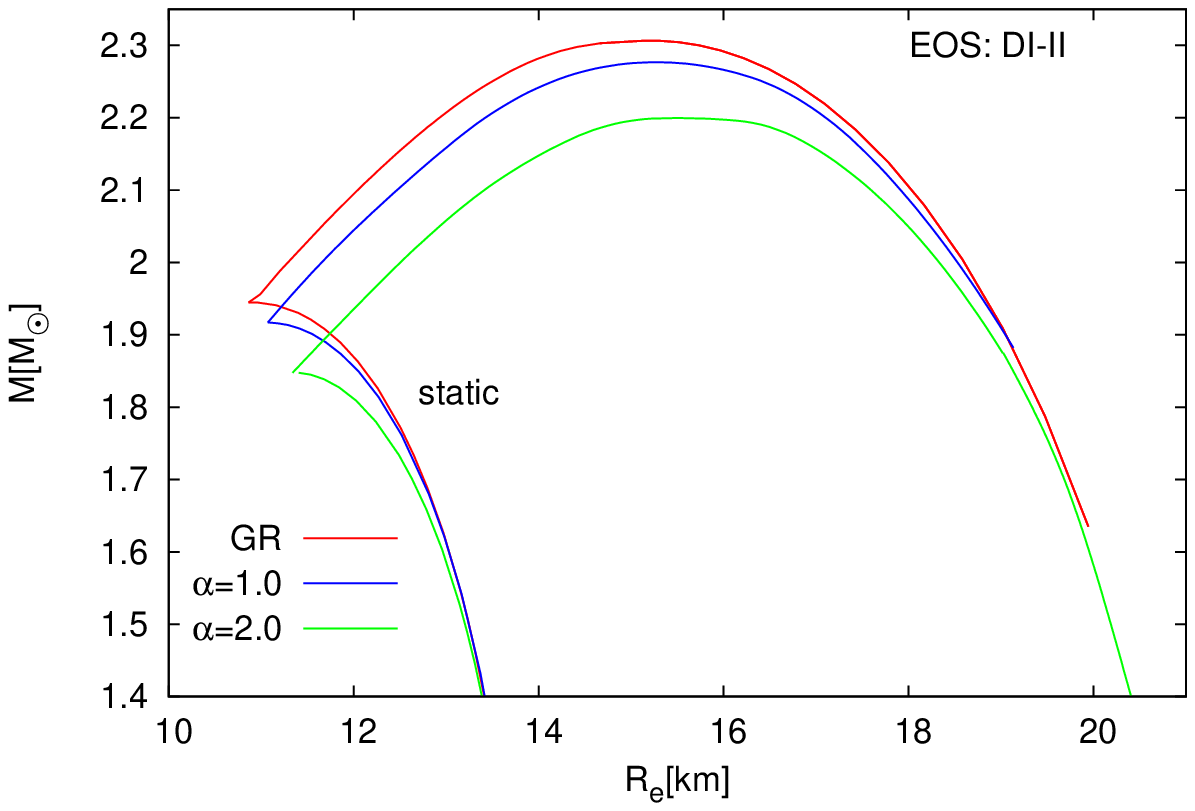}

\end{center}
  \vspace{-0.5cm}
\caption{(a) Mass-radius relation
for the physically relevant domain of neutron star models,
limited by the static solutions, the solutions at the Kepler limit
and the solutions at the secular instability line
for GB coupling constants $\alpha=0$, 1 and 2.
(a) EOS FPS, (b) EOS DI-II.
The mass $M$ is shown in units of $M_\odot$, the radius $R_e$ in km.
}
\label{fig1}
\end{figure}

Let us start by briefly recalling the mass-radius relation of the
neutron star models studied in \cite{Kleihaus:2016dui}.
The physically relevant domain is delimited by several
sequences of neutron star solutions, which comprise
(i) the static neutron stars up to the maximum mass,
(ii) the set of neutron stars along the secular instability line,
formed by the set of neutron stars with maximum mass at fixed angular momentum,
and (iii) the set of neutron stars rotating at the Kepler limit.
In the Kepler limit the angular velocity $\Omega_K$ is reached,
which corresponds to the angular velocity of a particle in geodesic motion
at the equatorial surface of the star.

The mass-radius relation for the above limiting sequences of
neutron star models for the two EOSs FPS and EOS DI-II and three values 
of the GB coupling constant, $\alpha=0$, 1 and 2,
is exhibited in Fig.~\ref{fig1}.
The value $\alpha=0$ corresponds to the case of GR, which features
the largest domain of existence of physically relevant models.
With increasing GB coupling constant the domain and the associated
maximal mass decrease.

\begin{figure}[p!]
\begin{center}
\mbox{\hspace{0.2cm}
{\hspace{-1.0cm}
\includegraphics[height=.25\textheight, angle =0]{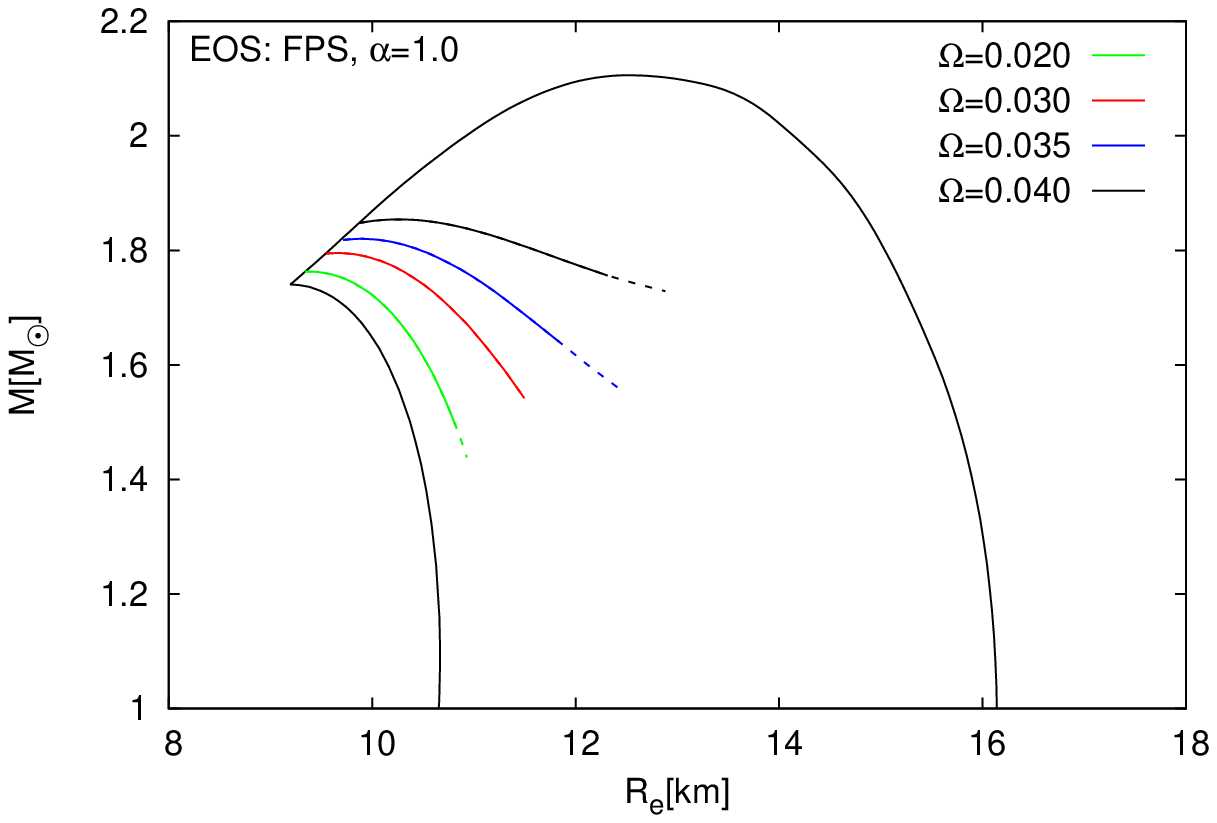}
}
{\hspace{-0.5cm}
\includegraphics[height=.25\textheight, angle =0]{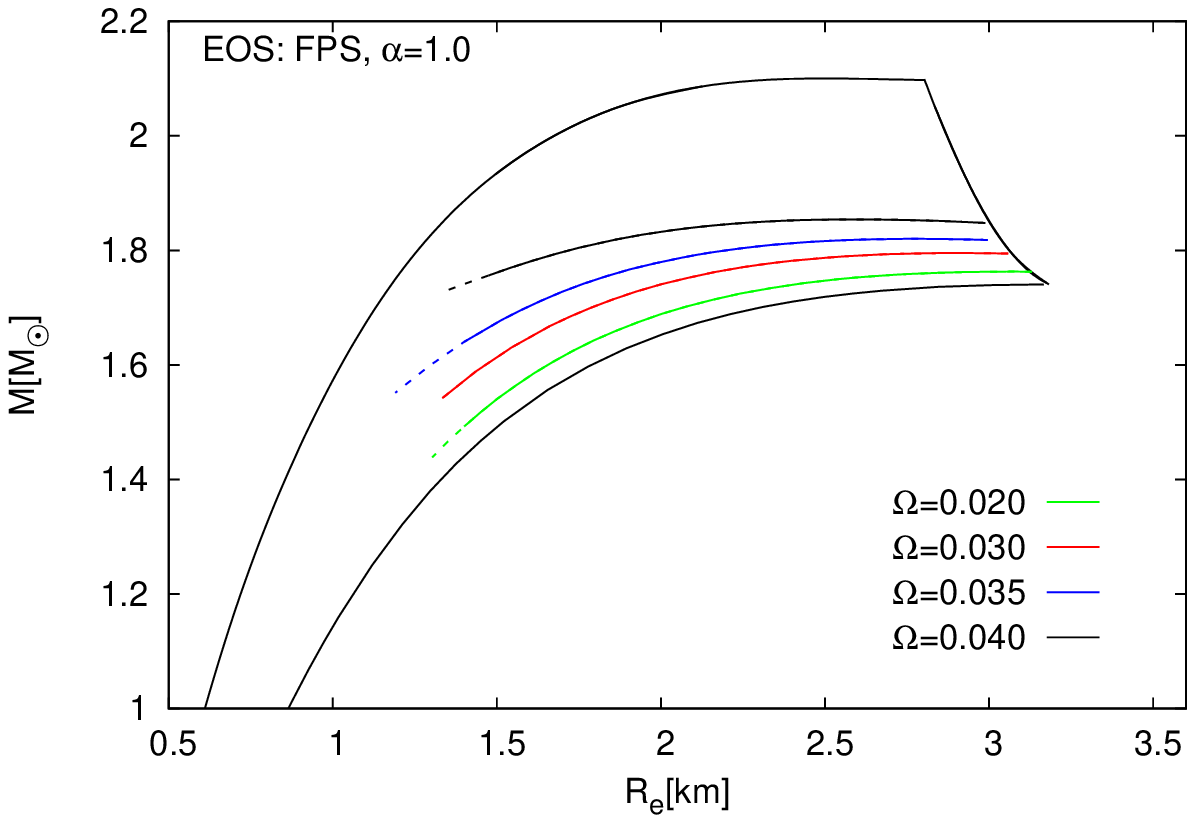}
}
}
\mbox{\hspace{0.2cm}
{\hspace{-1.0cm}
\includegraphics[height=.25\textheight, angle =0]{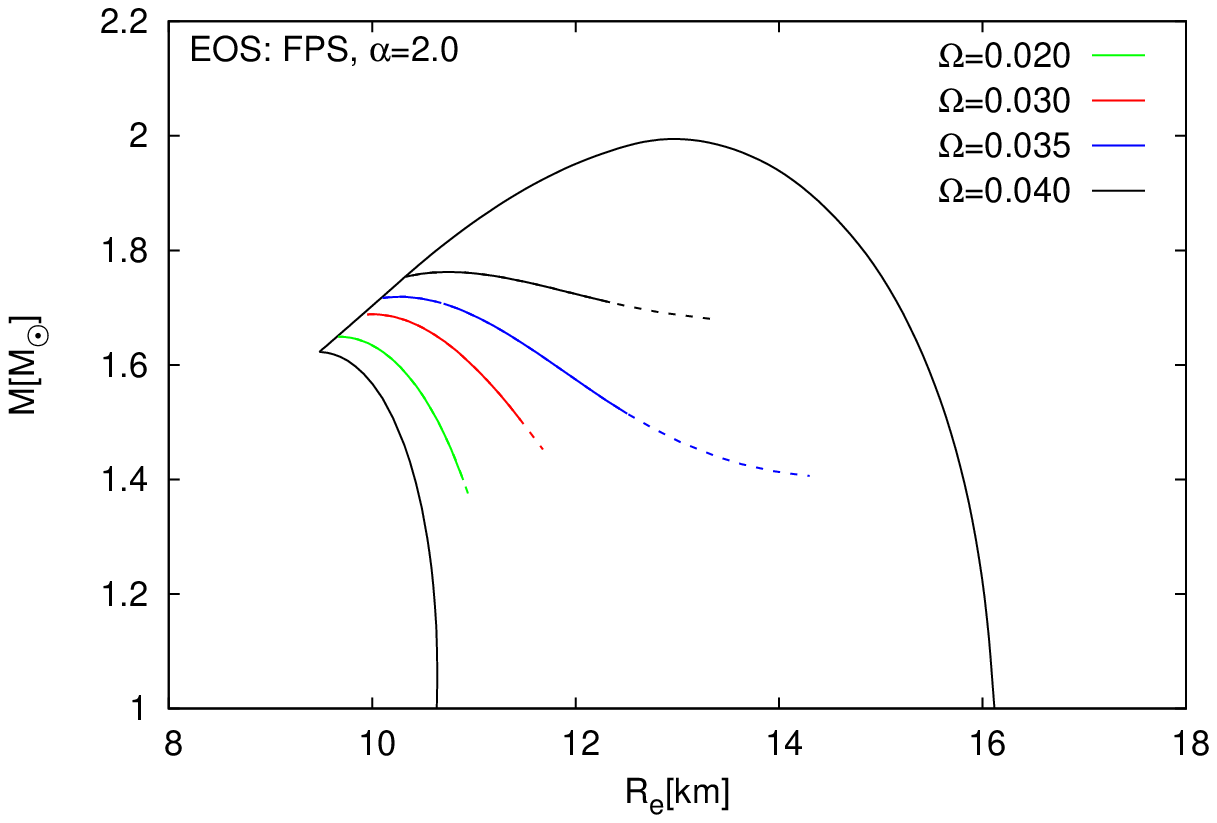}
}
{\hspace{-0.5cm}
\includegraphics[height=.25\textheight, angle =0]{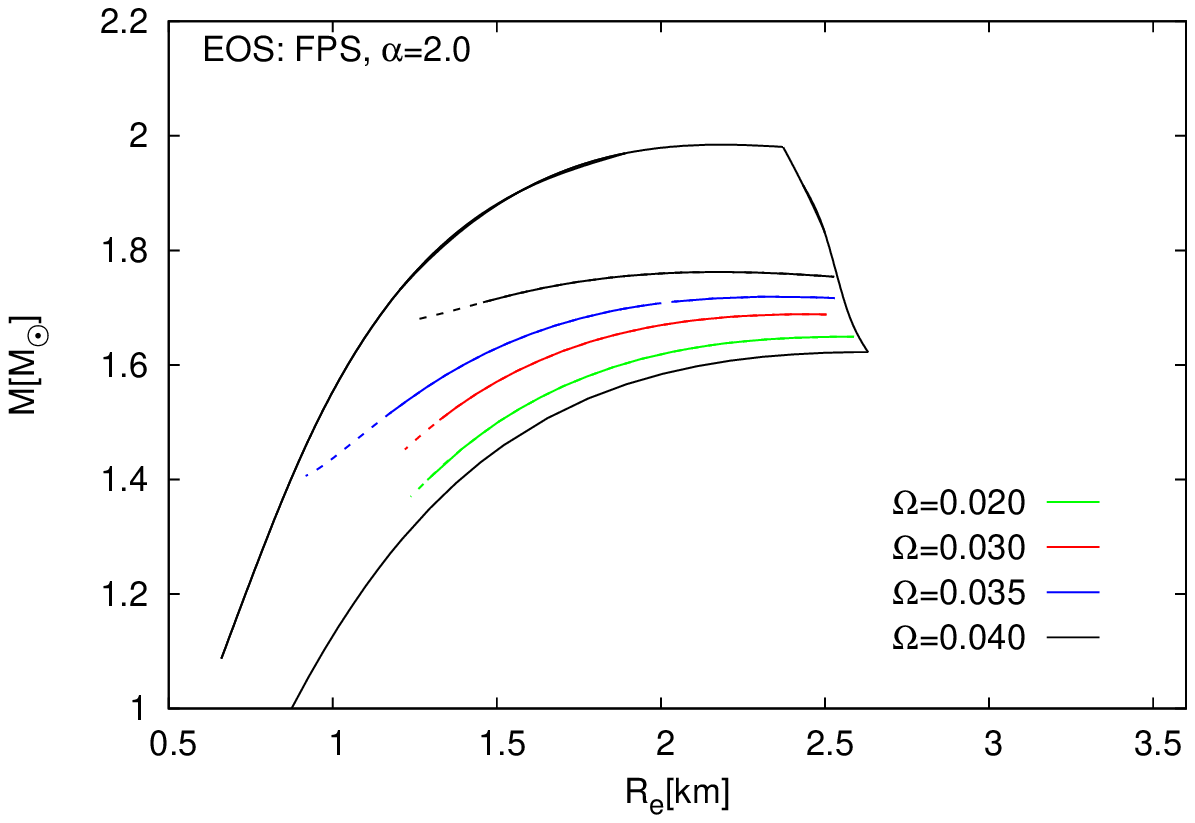}
}
}
\mbox{\hspace{0.2cm}
{\hspace{-1.0cm}
\includegraphics[height=.25\textheight, angle =0]{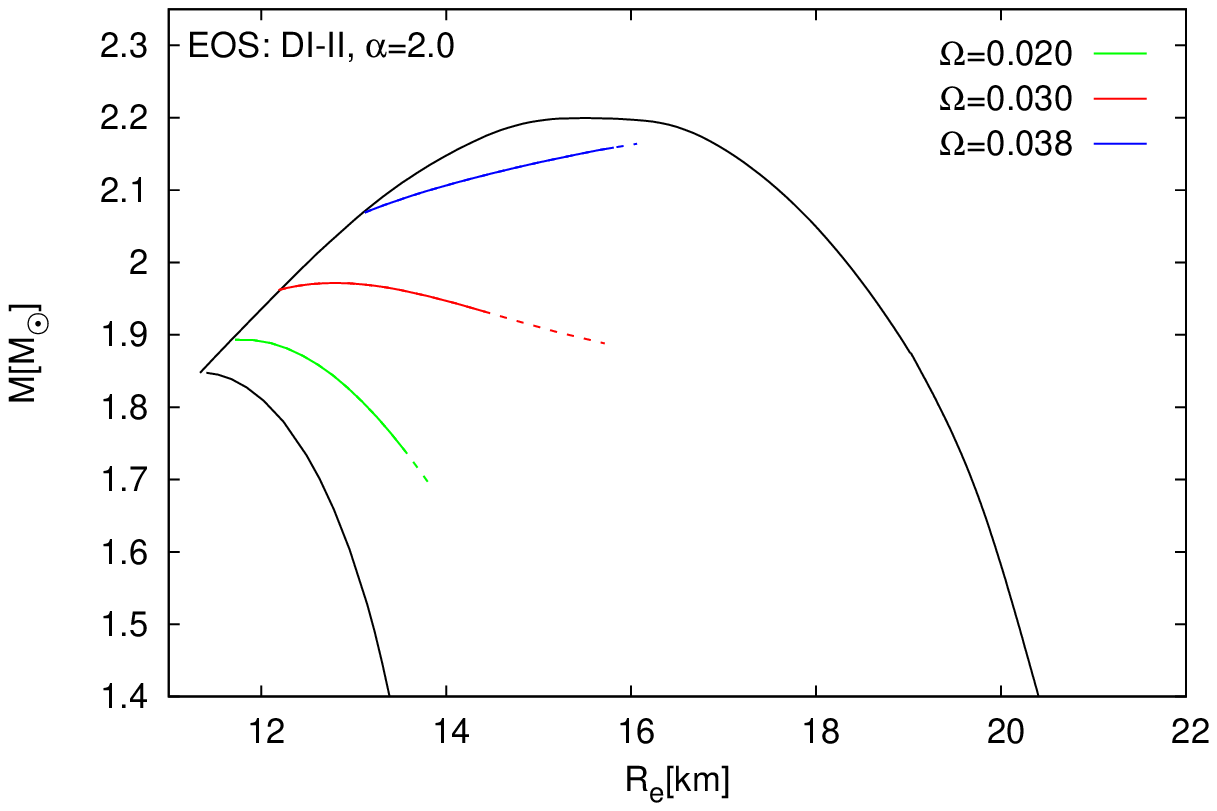}
}
{\hspace{-0.5cm}
\includegraphics[height=.25\textheight, angle =0]{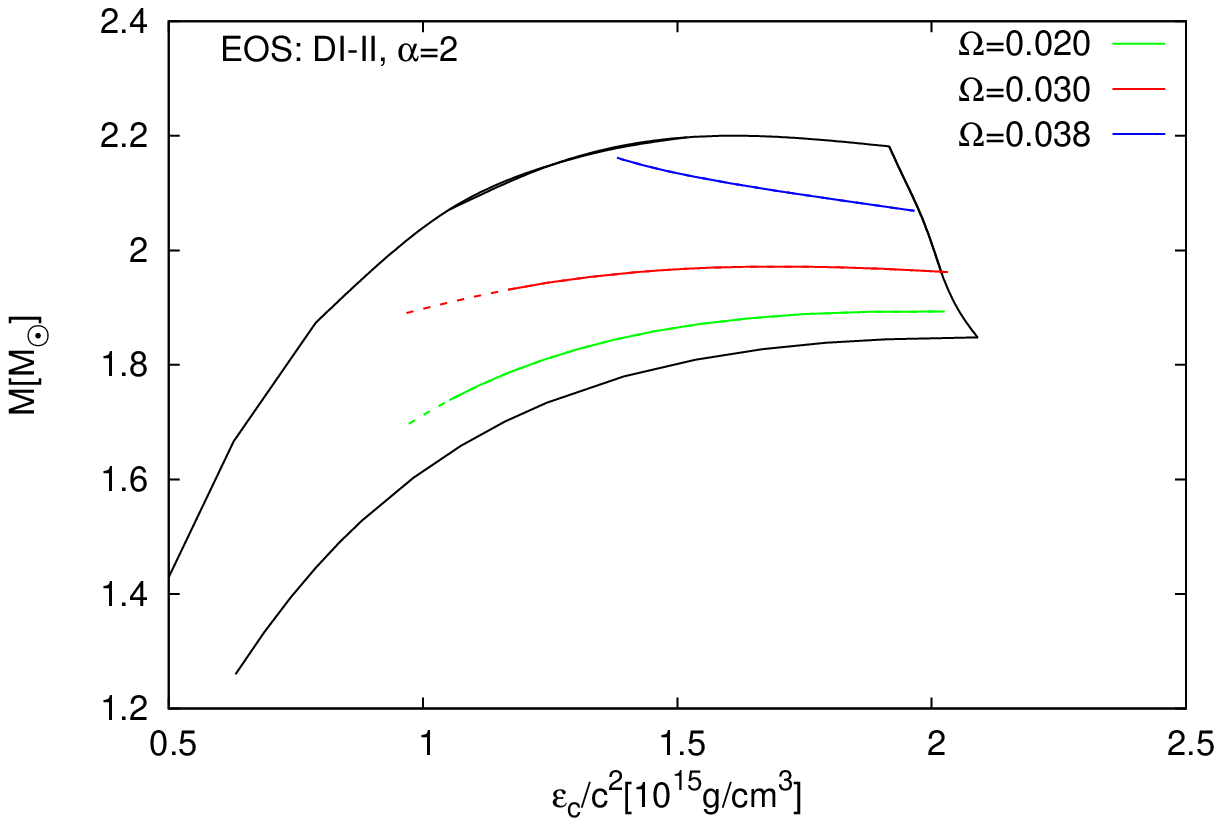}
}
}
\end{center}
\vspace{-0.5cm}
\caption{Range of the existence of ISCOs for coupling $\beta=0$
for fixed angular velocities $\Omega$ of neutron star models 
for the EOSs FPS ($\alpha=1$, 2) and EOS DI-II ($\alpha=2$).
Solid lines indicate solid numerical evidence,
dashed lines uncertain (see text) numerical evidence.
The range is indicated in the corresponding
mass-radius diagram in the left set of figures
and in the mass-energy density diagram in the right set of figures.
}
\label{Fig2}
\end{figure}

Evidently, ISCOs should reside outside the neutron stars.
Thus the neutron star surface represents an inner boundary
for the existence of ISCOs. 
ISCOs then exist for neutron star models only above a minimum mass.
In GR the dependence of this minimum mass on the angular velocity
of the stars has been studied for various EOS 
\cite{Cipolletta:2016yqv,Luk:2018xmt}.
Here we address this minimum mass in dEGB theory.

For that purpose
we exhibit in Fig.~\ref{Fig2} again the physically relevant domain
in mass-radius diagrams (left column), but now also in mass-energy density
diagrams (right column), 
for the EOSs FPS ($\alpha=1$: upper panels, $\alpha=2$: middle panels) 
and EOS DI-II ($\alpha=2$: lower panels).
The colored lines inside these domains indicate the presence of ISCOs
for sets of neutron star models with fixed rotational velocity
$\Omega$ and for coupling parameter $\beta=0$.
As long as these colored lines are solid, the numerical evidence for
their existence is solid. 
However, when the numerical accuracy for the effective potential 
is no longer very high,
it is hard to decide, whether an ISCOs still exists.
The numerical evidence then becomes less clear, and 
we have denoted this uncertain range with dashed lines.

\begin{figure}[h!]
\begin{center}
\mbox{\hspace{0.2cm}
{\hspace{-1.0cm}
\includegraphics[height=.25\textheight, angle =0]{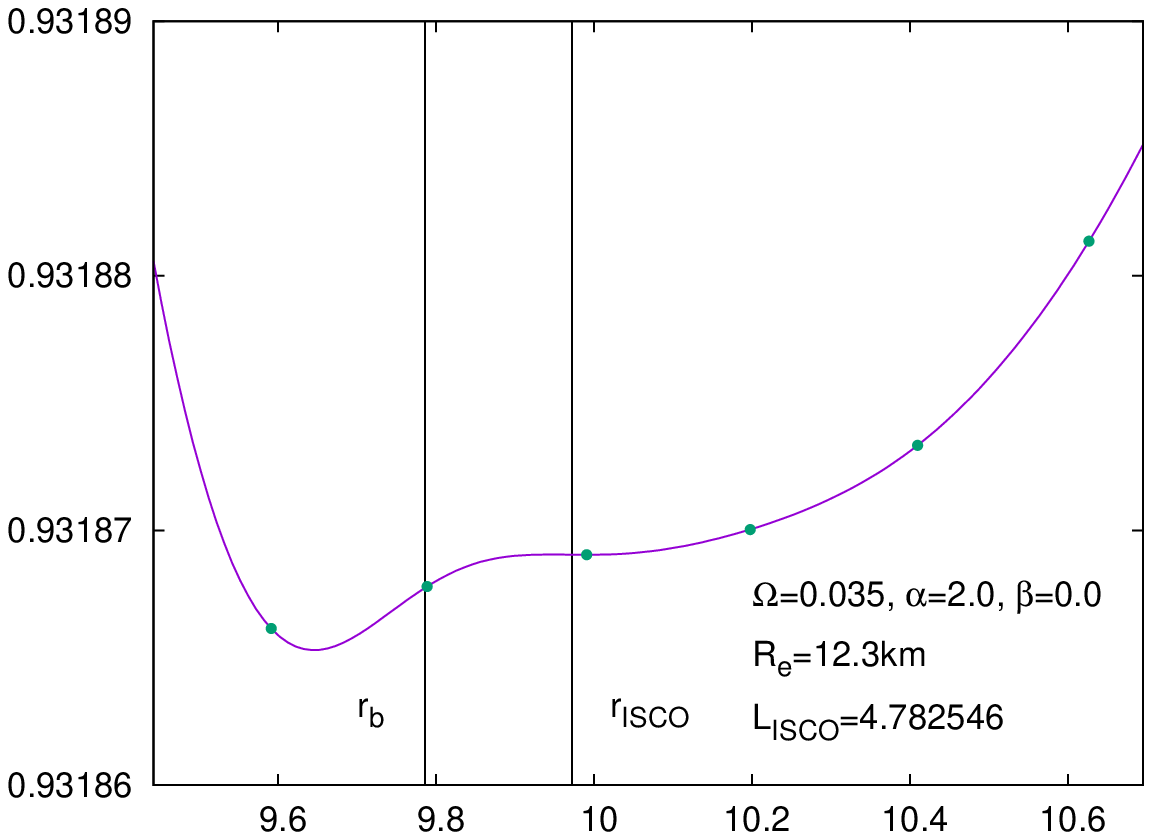}
}
{\hspace{-0.5cm}
\includegraphics[height=.25\textheight, angle =0]{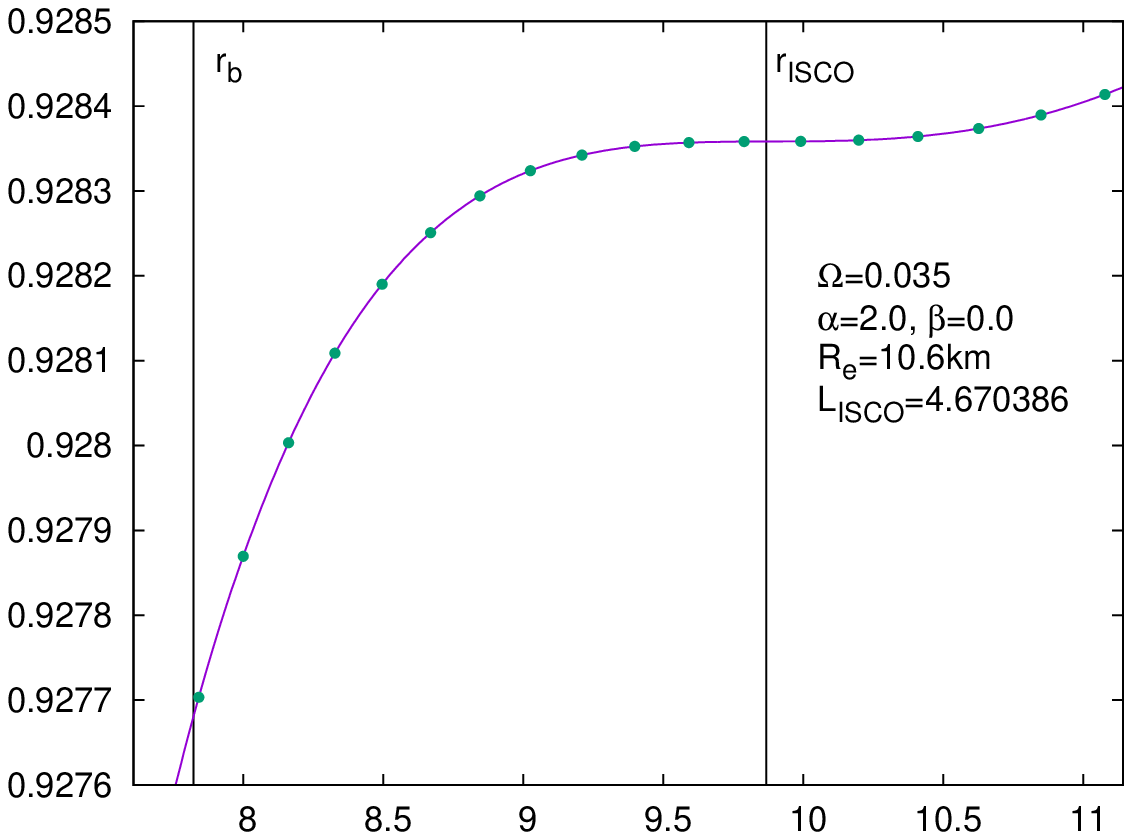}
}
}
\end{center}
\vspace{-0.5cm}
\caption{Effective potential $V_+$ for 
angular velocity $\Omega=0.035$ of the neutron star model,
EOSs FPS, GB coupling constant $\alpha=2$,
matter coupling $\beta=0$,
and equatorial radius of the star $R_{\rm e}=12.3$ km (left),
and $R_{\rm e}=10.6$ km (right).
Indicated are the radial surface coordinate of the neutron star $r_b$
and the coordinate location of the (tentative) ISCO $r_{\rm ISCO}$.
}
\label{Fig3}
\end{figure}

As an example for such a less clear cut case, we exhibit in 
Fig.~\ref{Fig3} (left) the effective potential $V_+$ for a rapidly
rotating neutron star model with $\Omega=0.035$ 
for the EOSs FPS, GB coupling $\alpha=2$, matter coupling
$\beta=0.0$, and ISCO angular momentum $L=L_{\rm ISCO}$.
The dots indicate the mesh points where the solution is computed.
The solid curve corresponds to the interpolated values. We note that
the situation is similar for $\beta=0$ and $\beta=0.5$.
For comparison Fig.~\ref{Fig3} (right) shows the effective potential $V_+$
for $L=L_{\rm ISCO}$ in case the ISCO is located at some larged distance
from the star boundary,
where the numerical accuracy is much better.

As a general trend we extract from theses calculations,
that the higher the angular velocity of the star, the
higher the minimum mass needed for ISCOs to exist.
This trend seen in dEGB models is following the one seen in GR models.
But just as the overall maximum mass decreases with 
increasing dEGB coupling $\alpha$, the minimum mass
decreases with increasing $\alpha$, as well.

\subsection{ISCO radii and frequencies}

The ISCO has been studied in GR for rotating neutron stars
subject to many diffent EOS
(see e.g., \cite{Miller:1998gr,Shibata:1998xw,Zdunik:2000qn,
GondekRosinska:2000mp,Pachon:2006an,Bhattacharyya:2011wm,
Pappas:2012nv,Pappas:2012nt, Gondek-Rosinska:2014aaa,
Torok:2014ina,Cipolletta:2016yqv,Luk:2018xmt}).
Let us now consider the ISCOs in dEGB theory
in some detail and their dependence
on the GB coupling constant $\alpha$,
employing the above sets of 
neutron star models \cite{Kleihaus:2016dui}.

In particular, let us consider sets for fixed EOS and fixed $\alpha$,
and either i) fixed scaled angular momentum $j=J/M^2$ of the star,
or ii) fixed angular velocity $\Omega$ of the star.
Let us note, that $\Omega=0.01$ corresponds to a rotation frequency of 323 Hz,
while the fastest spinning known neutron star has a frequency of 716 Hz
\cite{Hessels:2006ze}, i.e., $\Omega \approx 0.022$.

\begin{figure}[h!]
\begin{center}
\mbox{\hspace{0.2cm}
{\hspace{-1.0cm}
\includegraphics[height=.25\textheight, angle =0]{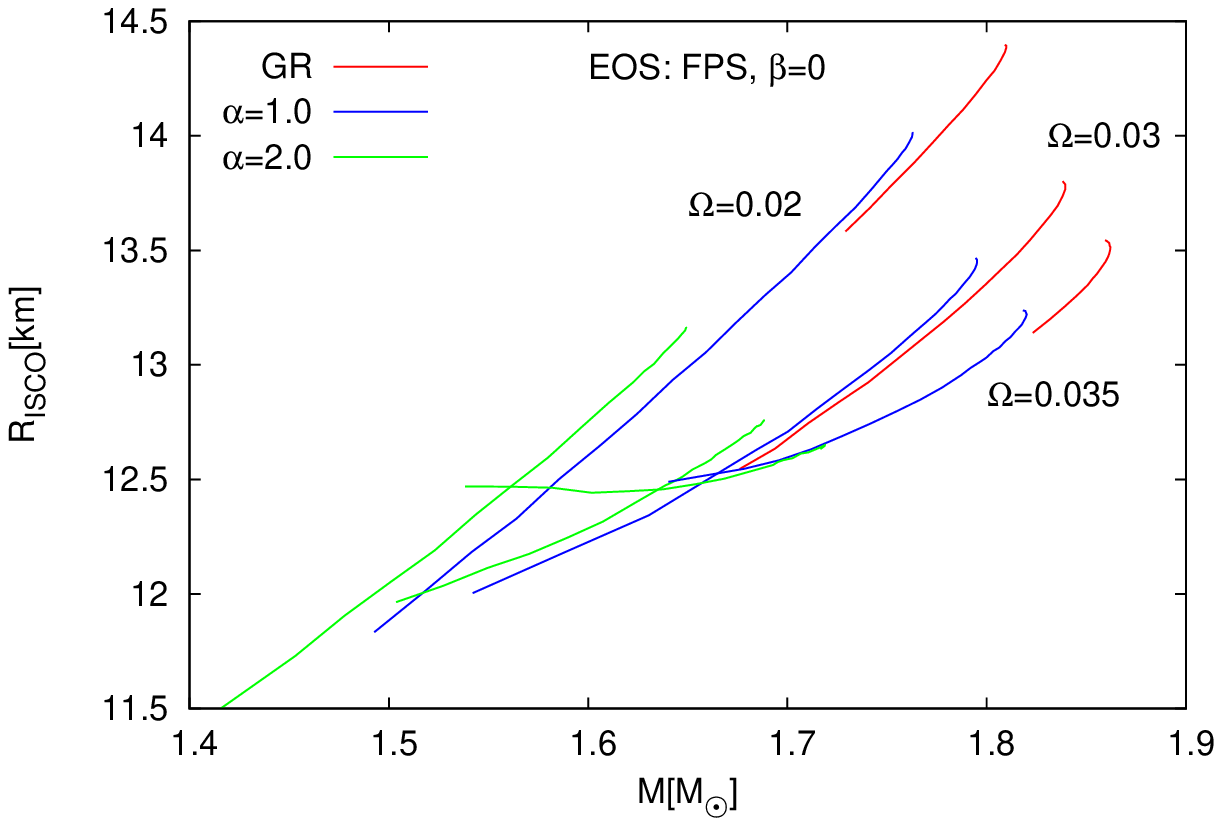}
}
{\hspace{-0.5cm}
\includegraphics[height=.25\textheight, angle =0]{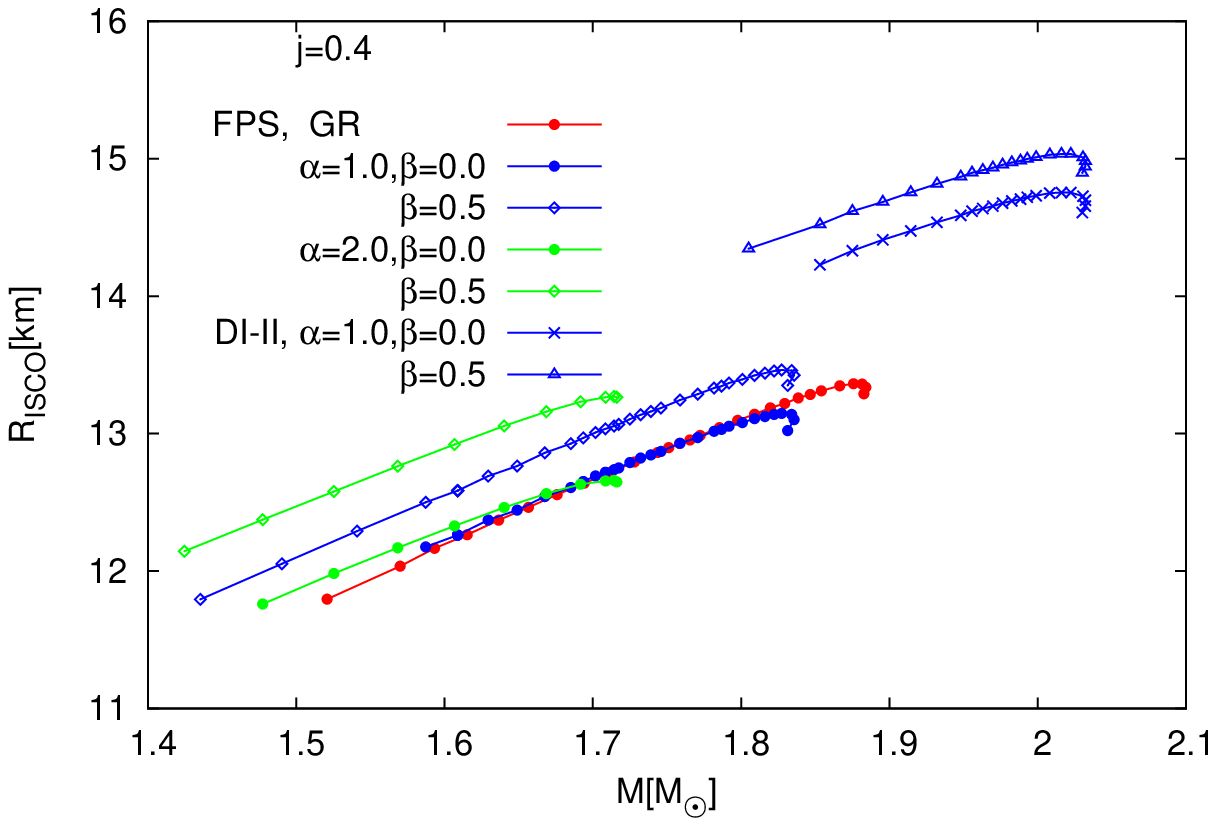}
}
}
\end{center}
\vspace{-0.5cm}
\caption{ISCO radius $R_{\rm ISCO}$ in km versus the neutron star mass
$M$ in solar masses $M_\odot$.
Left: Fixed values of the angular velocity $\Omega$ of the star,
$\Omega=0.02$, $0.03$ and $0.035$, matter coupling $\beta=0$,
EOS FPS, for GR (red), dEGB with $\alpha=1.0$ (blue) and
dEGB with $\alpha=2.0$ (green).
Right: fixed value of the scaled angular momentum $j=0.4$ of the star,
matter coupling $\beta=0$ (dots) and $\beta=0.5$ (diamonds),
EOS FPS, for GR (red), dEGB with $\alpha=1.0$ (blue) and
dEGB with $\alpha=2.0$ (green),
as well as EOS DI-II, matter coupling $\beta=0$ (crosses) 
and $\beta=0.5$ (triangles) 
for dEGB with $\alpha=1.0$ (blue).
}
\label{Fig4}
\end{figure}

In Fig.\ref{Fig4} we exhibit the radius of the ISCO $R_{\rm ISCO}$
in km versus the mass of the neutron star $M$ in units of the solar mass
$M_\odot$ for several families of rapidly rotation neutron star models.
Fig.\ref{Fig4} (left) exhibits results for the EOS FPS, where
the GB coupling $\alpha$ is fixed to zero, i.e., reproducing GR, 
to $\alpha=1$ and $\alpha=2$. Also fixed are 
the value of the matter coupling $\beta=0$,
and the angular velocity of the stars,
which assumes values $\Omega=0.02$, 0.03 and 0.035.
As seen in the figure, the different values of $\Omega$ lead to shifts
towards smaller ISCO radii for given GB values. Likewise, an increase
in the GB coupling constant leads to smaller ISCO radii for a given
$\Omega$.

Fig.\ref{Fig4} (right) on the other hand, considers 
families of neutron star models for a fixed value
of the scaled angular momentum $j=0.4$. 
For the EOS FPS the GB coupling $\alpha$ is fixed to zero, 
i.e., reproducing GR, to $\alpha=1$ and $\alpha=2$,
while the matter coupling assumes the values $\beta=0$ and $\beta=0.5$.
In addition, the families for $\alpha=1$
and $\beta=0$ and $\beta=0.5$ are shown for the EOS DI-II.
The figure shows, that the increase of $\beta$ leads to an
increase of the ISCO radii, that rises strongly with the increase
of $\alpha$.
Moreover, the figure demonstrates the strong dependence of the
ISCO radius on the EOS.

\begin{figure}[h!]
\begin{center}
\mbox{\hspace{0.2cm}
{\hspace{-1.0cm}
\includegraphics[height=.25\textheight, angle =0]{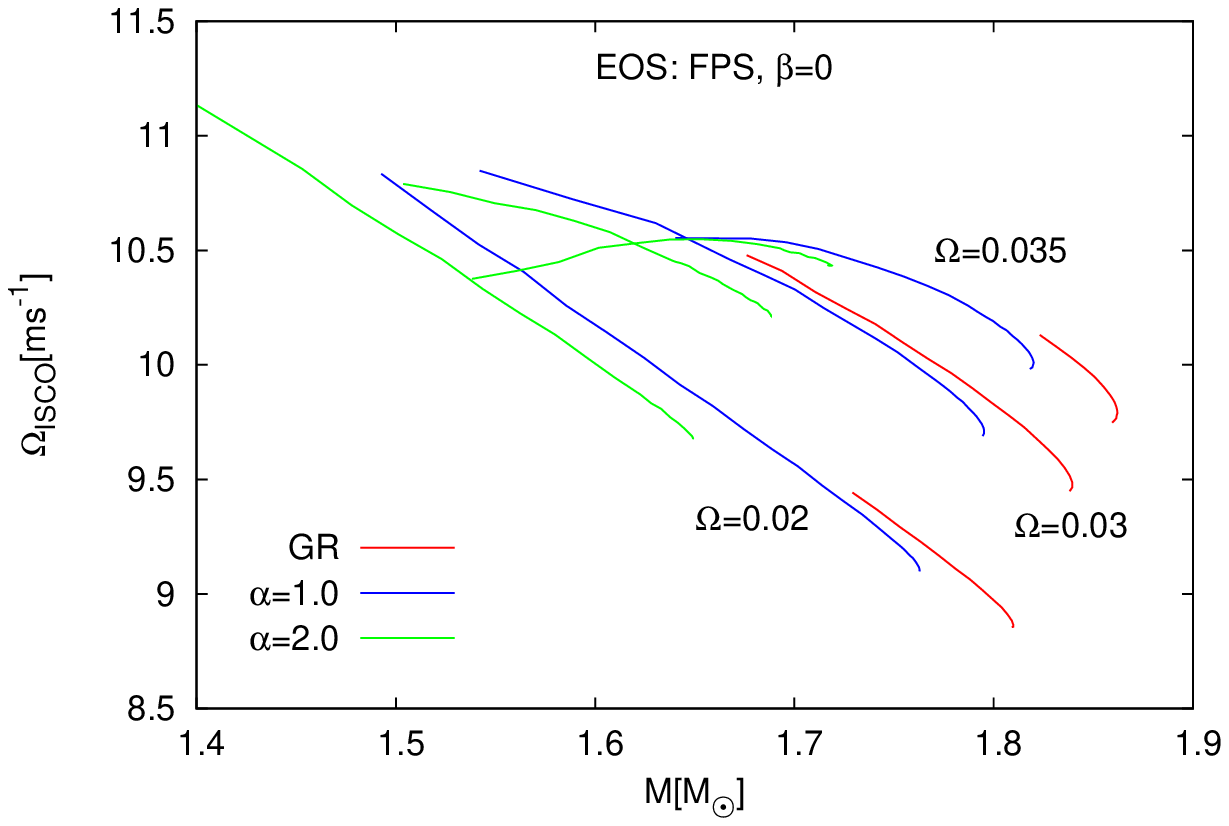}
}
{\hspace{-0.5cm}
\includegraphics[height=.25\textheight, angle =0]{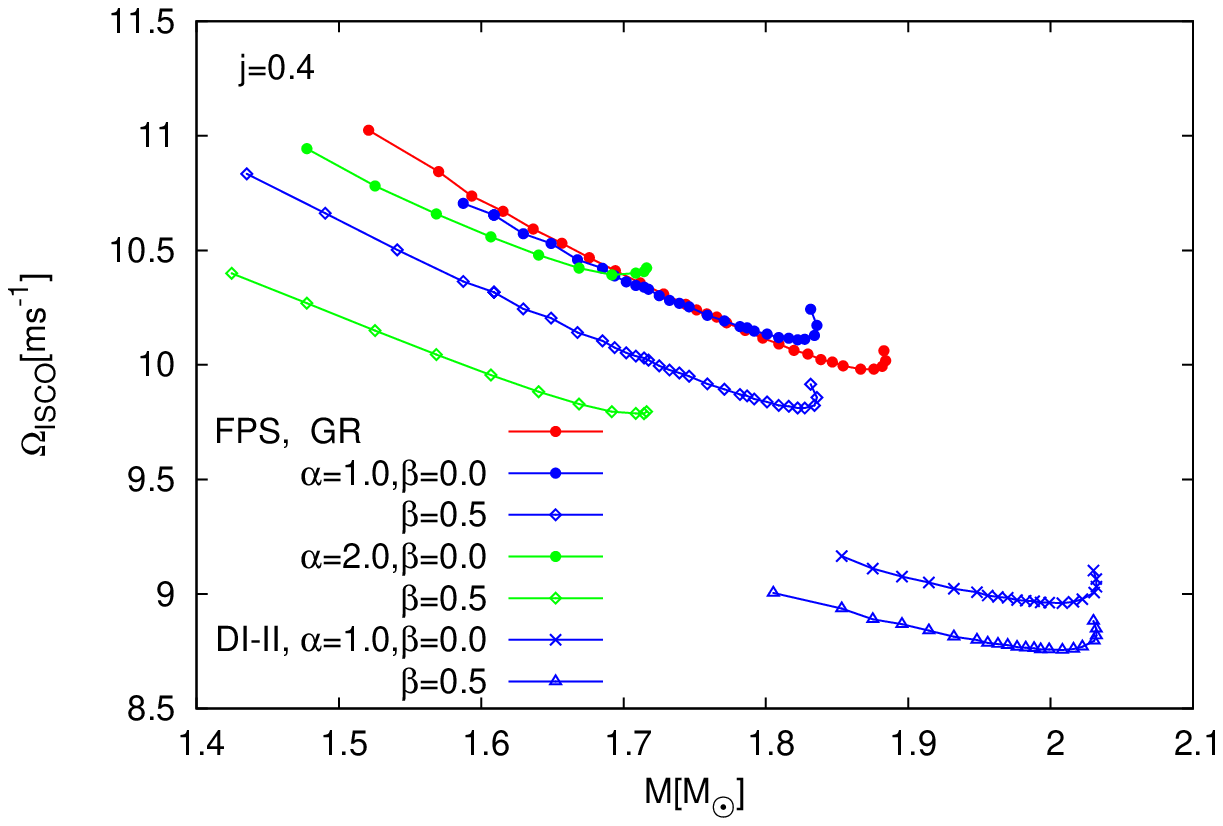}
}
}
\end{center}
\vspace{-0.5cm}
\caption{ISCO angular velocity $\Omega_{\rm ISCO}$ in ms$^{-1}$ 
versus the neutron star mass
$M$ in solar masses $M_\odot$.
Left: Fixed values of the angular velocity $\Omega$ of the star,
$\Omega=0.02$, $0.03$ and $0.035$, matter coupling $\beta=0$,
EOS FPS, for GR (red), dEGB with $\alpha=1.0$ (blue) and
dEGB with $\alpha=2.0$ (green).
Right: fixed value of the scaled angular momentum $j=0.4$ of the star,
matter coupling $\beta=0$ (dots) and $\beta=0.5$ (diamonds),
EOS FPS, for GR (red), dEGB with $\alpha=1.0$ (blue) and
dEGB with $\alpha=2.0$ (green),
as well as EOS DI-II, for dEGB,
matter coupling $\beta=0$ (crosses) and $\beta=0.5$ (triangles) 
with $\alpha=1.0$ (blue).
}
\label{Fig5}
\end{figure}

In Fig.~\ref{Fig5} we exhibit the ISCO angular velocity 
$\Omega_{\rm ISCO}$ in ms$^{-1}$ 
versus the mass of the neutron star $M$ in units of the solar mass
$M_\odot$ for the same families of rapidly rotating neutron star models
as in Fig.~\ref{Fig4}.
As seen in Fig~\ref{Fig5} (left), 
the different values of the angular velocity $\Omega$ of the star 
now lead to shifts towards smaller ISCO angular velocities for given GB values, 
while an increase in the GB coupling constant 
leads to smaller ISCO angular velocities 
for a given angular velocity $\Omega$ of the star.
Fig.~\ref{Fig5} (right) shows, that the increase of 
the matter coupling $\beta$ now leads to a
decrease of the ISCO angular velocities, that grows with increasing
$\alpha$.
In addition, the figure demonstrates the strong dependence of the
ISCO  angular velocity on the EOS.

\begin{figure}[h!]
\begin{center}
\mbox{\hspace{0.2cm}
{\hspace{-1.0cm}
\includegraphics[height=.25\textheight, angle =0]{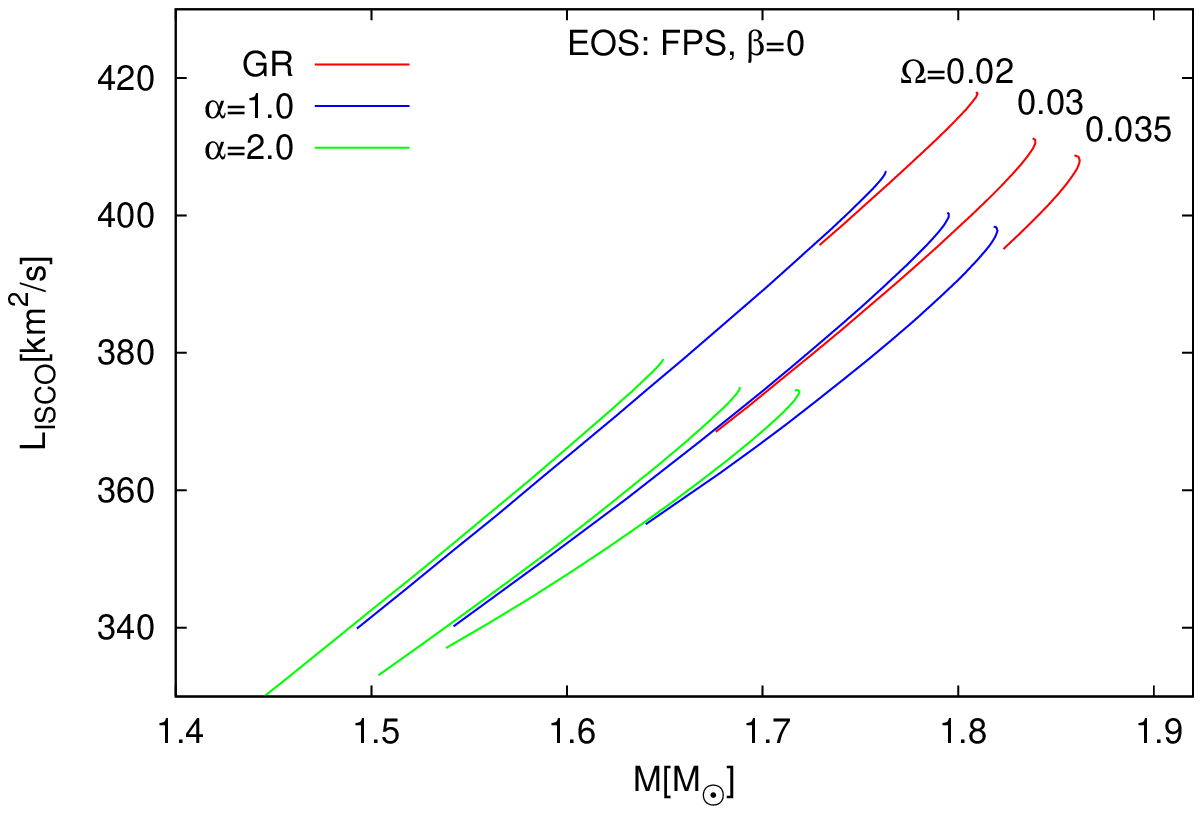}
}
{\hspace{-0.5cm}
\includegraphics[height=.25\textheight, angle =0]{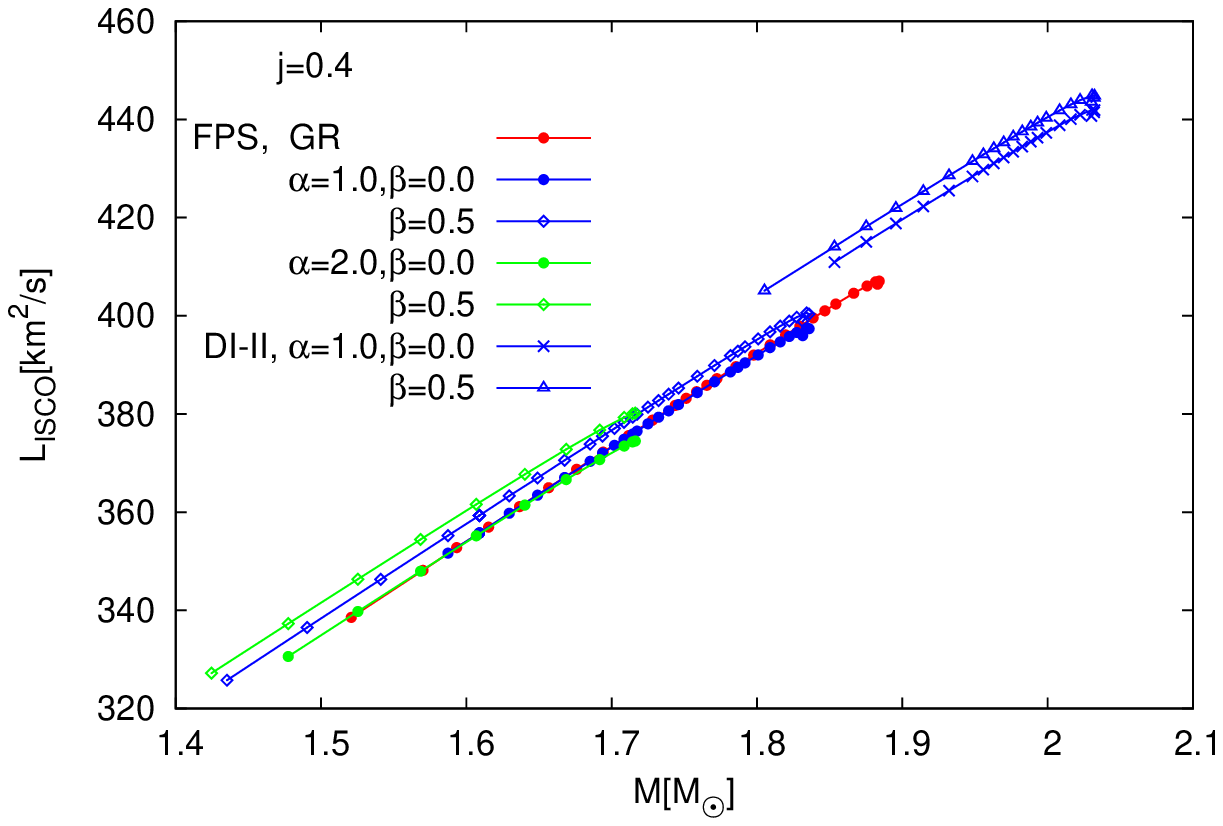}
}
}
\end{center}
\vspace{-0.5cm}
\caption{ISCO angular momentum $L_{\rm ISCO}$ 
in km$^2$s$^{-1}$ versus the neutron star mass
$M$ in solar masses $M_\odot$.
Left: Fixed values of the angular velocity $\Omega$ of the star,
$\Omega=0.02$, $0.03$ and $0.04$, matter coupling $\beta=0$,
EOS FPS, for GR (red), dEGB with $\alpha=1.0$ (blue) and
dEGB with $\alpha=2.0$ (green).
Right: fixed value of the scaled angular momentum $j=0.4$ of the star,
matter coupling $\beta=0$ (dots) and $\beta=0.5$ (diamonds),
EOS FPS, for GR (blue), dEGB with $\alpha=1.0$ (red) and
dEGB with $\alpha=2.0$ (green),
as well as EOS DI-II, matter coupling $\beta=0$ (crosses) 
and $\beta=0.5$ (triangles) 
for dEGB with $\alpha=1.0$ (blue).
}
\label{Fig6}
\end{figure}

In Fig.~\ref{Fig6} we show the analogous sets of curves for the
ISCO angular momentum $L_{\rm ISCO}$ in km$^2$s$^{-1}$ versus the 
neutron star mass. Here the dependencies on the parameters
and EOS are also present but weaker than in the previous two cases,
making the ISCO angular momentum $L_{\rm ISCO}$ a good new
candidate when looking for universal relations.

\subsection{Universal relations}

Universal relations represent a unique means to learn about
the EOS and the theory of gravity
\cite{Yagi:2016bkt,Doneva:2017jop}.
In these universal relations one typically considers 
appropriately scaled physical properties, 
representing dimensionless quantities,
and studies their various relations.
For the scaling itself then the mass, radius or frequency of the
neutron stars are invoked with appropriate powers.

When the $I$-Love-$Q$ relations were generalized for rapidly rotating
neutron stars, it was realized, that the $I$-Love-$Q$ relations
did not yield single (best fit) curves but full surfaces
\cite{Doneva:2013rha,Pappas:2013naa,Chakrabarti:2013tca}.
Here the additional dimensionless physical quantity could be 
represented by the scaled angular momentum $j$ or the scaled
frequency in the form $\Omega M$ or $\Omega R$, but in each case 
corresponding to an additional dependence on the
rotation of the neutron star.

Recently an analysis of ISCO properties by Luk and Lin \cite{Luk:2018xmt}
has led to a set of new universal relations in GR.
There it was observed, that when $R_{\rm ISCO} \Omega$, 
i.e., the ISCO radius scaled by the star angular velocity,
is considered as a function of $M \Omega$,
i.e., the mass of the star scaled by its angular velocity,
the data points from 12 different EOS fall more of less on a single curve,
with deviations from this fitted universal curve
on the order of up to 6\%.
Similarly, when instead $\Omega/\Omega_{\rm ISCO}$,
i.e., the scaled ISCO angular velocity, is considered,
the data points from the same set of EOS yield a fitted
universal curve with deviations of the same order.

These universal relations of Luk and Lin \cite{Luk:2018xmt} 
(in terms of the frequency $f=\Omega/2\pi$)
correspond to
\begin{equation}
y = a_1 x + a_2 x^2 + a_3 x^3 + a_4 x^4 \ ,
\end{equation}
where for $y=R_{\rm ISCO} f$ and $x=M f$,
the fitting parameters are
$a_1=8.809,\, a_2=-9.166 \times 10^{-4},\,
a_3=8.787 \times 10^{-8},\,
a_4=-6.019 \times 10^{-12}$,
while for $y=f/f_{\rm ISCO}$
the fitting parameters are given by
$b_1=4.497 \times 10^{-4},\, b_2=-6.130 \times 10^{-8},\,
b_3=4.527 \times 10^{-12},\,
b_4=-1.446 \times 10^{-16}$.
For these universal curves the relative deviations $(\hat y-y)/y$
of the data points $\hat y$ are then on the few percent level
for a very large range of stars, with
$0 \le M f \le 5000 M_\odot {\rm Hz}$.

\begin{figure}[h!]
\begin{center}
\mbox{\hspace{0.2cm}
{\hspace{-1.0cm}
\includegraphics[height=.25\textheight, angle =0]{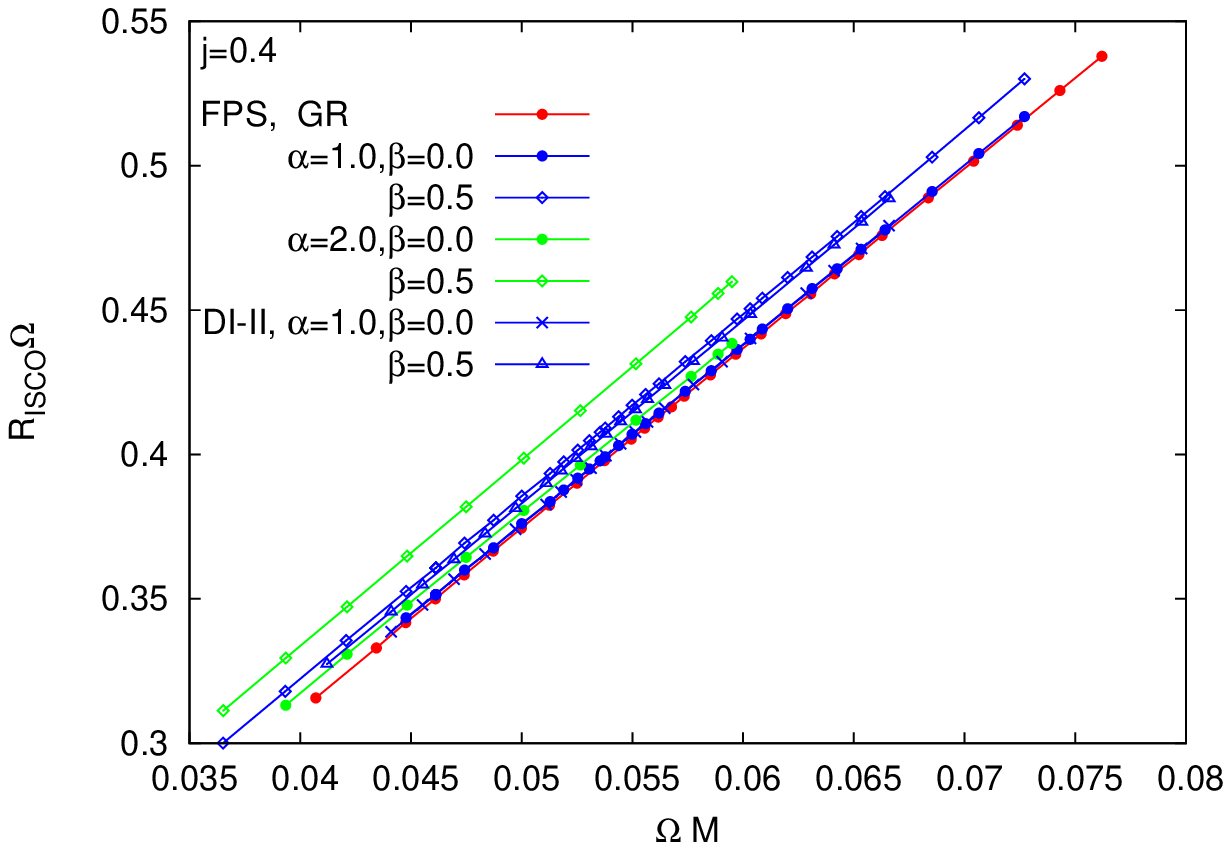}
}
{\hspace{-0.5cm}
\includegraphics[height=.25\textheight, angle =0]{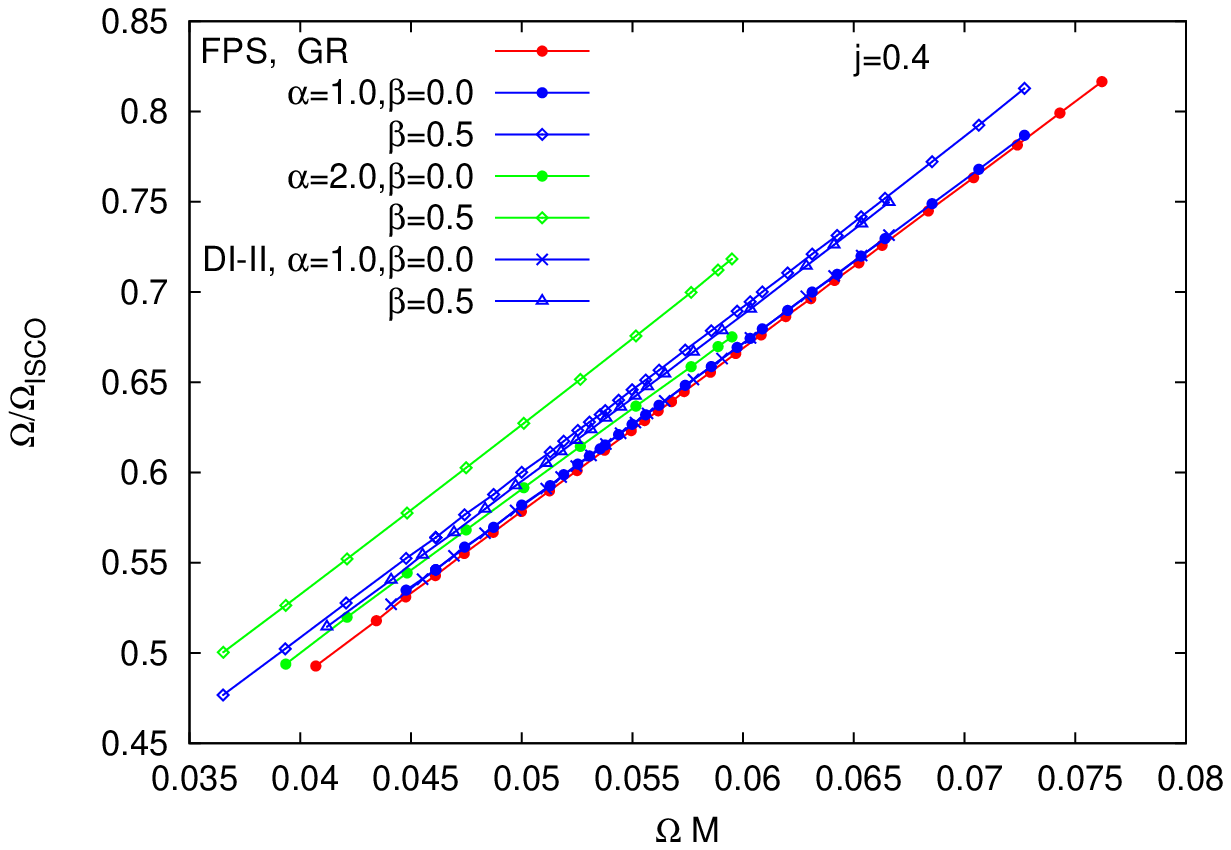}
}
}
\end{center}
\vspace{-0.5cm}
\caption{
Scaled ISCO radius $R_{\rm ISCO} \Omega$ (left)
and scaled ISCO angular velocity $\Omega/\Omega_{\rm ISCO}$ (right)
versus the scaled neutron star mass $M \Omega$.
Fixed value of the scaled angular momentum $j=0.4$ of the star,
matter coupling $\beta=0$ (dots) and $\beta=0.5$ (diamonds),
EOS FPS, for GR (red), dEGB with $\alpha=1.0$ (blue) and
dEGB with $\alpha=2.0$ (green),
as well as EOS DI-II, for dEGB with $\alpha=1.0$, $\beta=0$ (crosses),
and $\beta=0.5$ (triangles).
}
\label{Fig7}
\end{figure}

When comparing our GR data to these two fits, we find complete
agreement, with the two universal relations of \cite{Luk:2018xmt}
being satisfied within 1.5\%, respectively 2.5\%.
However, here our goal is to see the effects of the dEGB action
and the associated coupling constants $\alpha$ and $\beta$.
Let us therefore inspect what happens, as we consider
the data of Fig.~\ref{Fig4} (right) and Fig.~\ref{Fig5} (right),
appropriately scaled to match the above universal relations,
as exhibited in Fig.~\ref{Fig7}.

The figures show that universality is very well satisfied 
for GR and dEGB with $\alpha=1$ and $\beta=0$. 
Both satisfy more or less the same fits.
For $\beta=0.5$,
however, universality is slightly less well satisfied.
In particular, we notice that new fits should be made
for dEGB with $\alpha=1$ and $\beta=0.5$, since the
deviations from GR are getting larger, exceeding even
slightly the $\alpha=2$, $\beta=0$ deviations.
For $\alpha=2$ and $\beta=0.5$ the deviations from GR 
become even larger.
Thus observations of such large discrepancies between
the scaled ISCO radii or angular velocities might allow to
conclude, that a generalized theory of gravity might
be needed.

\begin{figure}[h!]
\begin{center}
\mbox{\hspace{0.2cm}
{\hspace{-1.0cm}
\includegraphics[height=.25\textheight, angle =0]{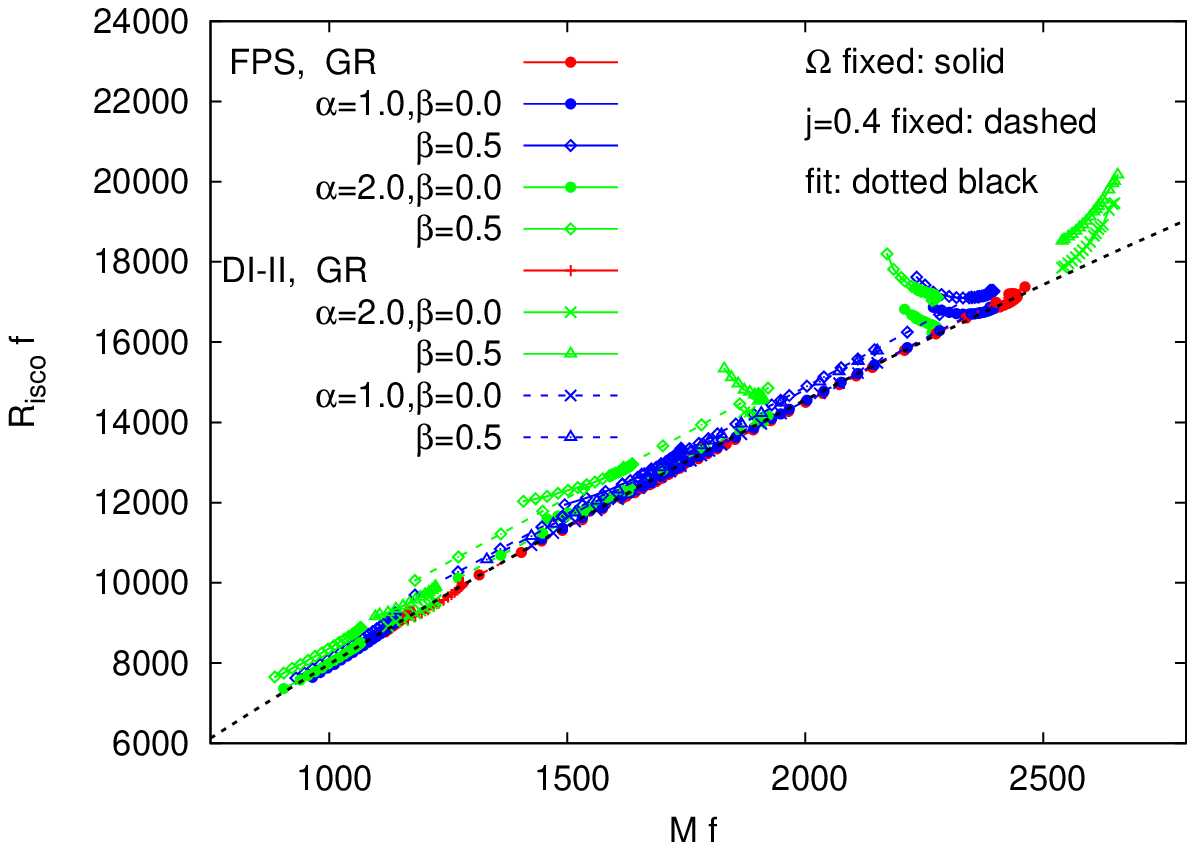}
}
{\hspace{-0.5cm}
\includegraphics[height=.25\textheight, angle =0]{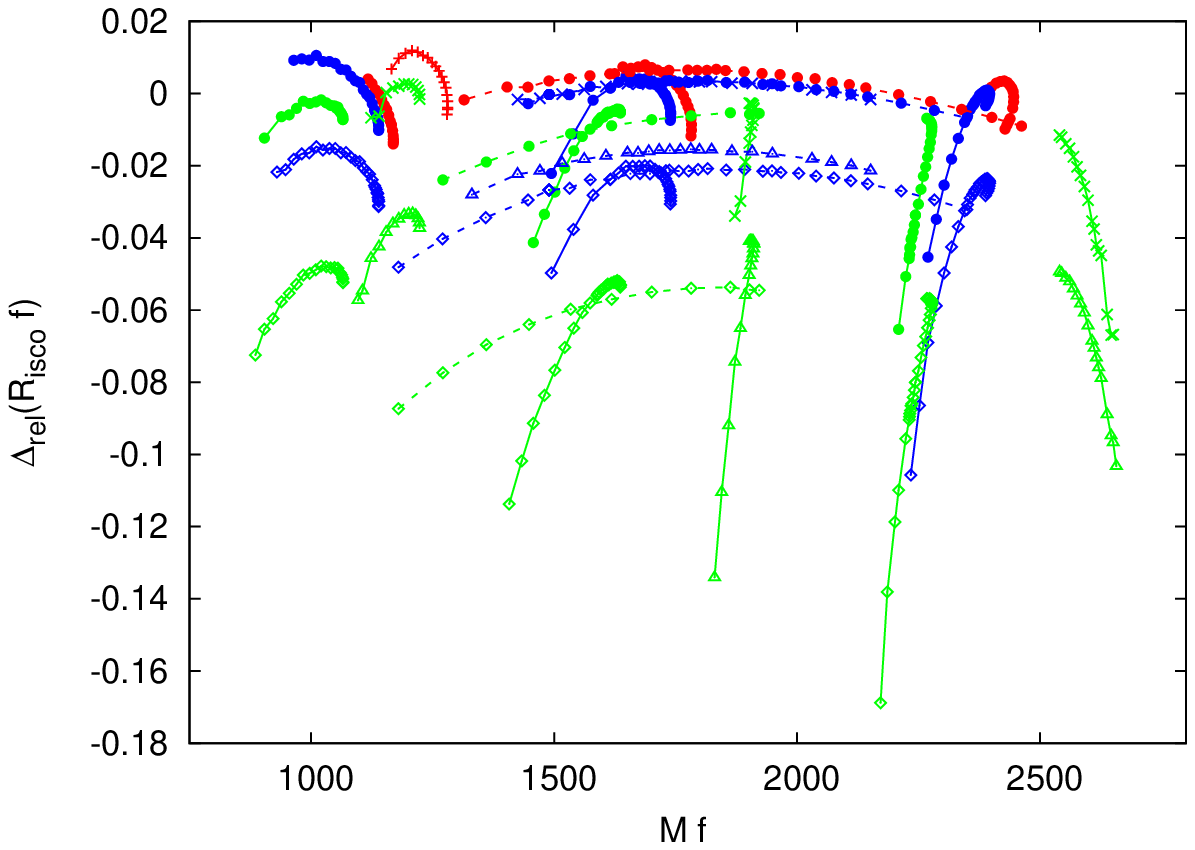}
}
}
\mbox{\hspace{0.2cm}
{\hspace{-1.0cm}
\includegraphics[height=.25\textheight, angle =0]{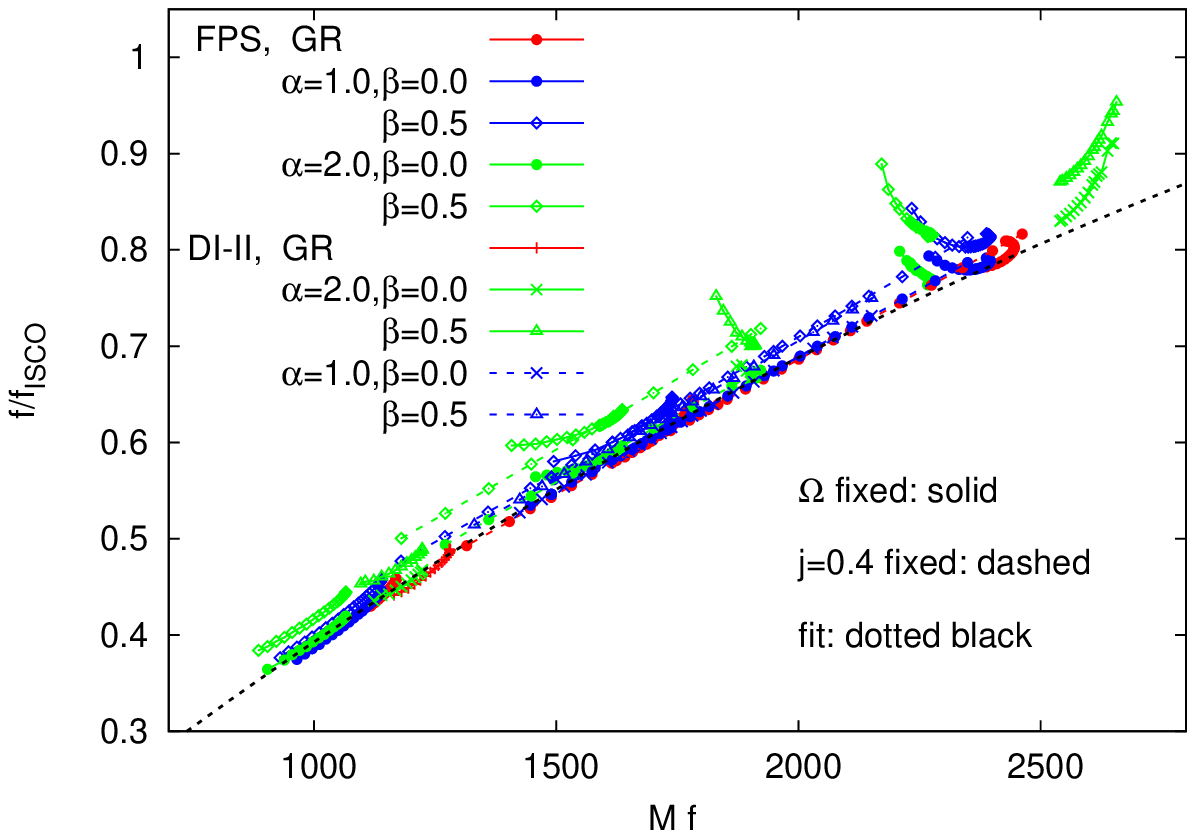}
}
{\hspace{-0.5cm}
\includegraphics[height=.25\textheight, angle =0]{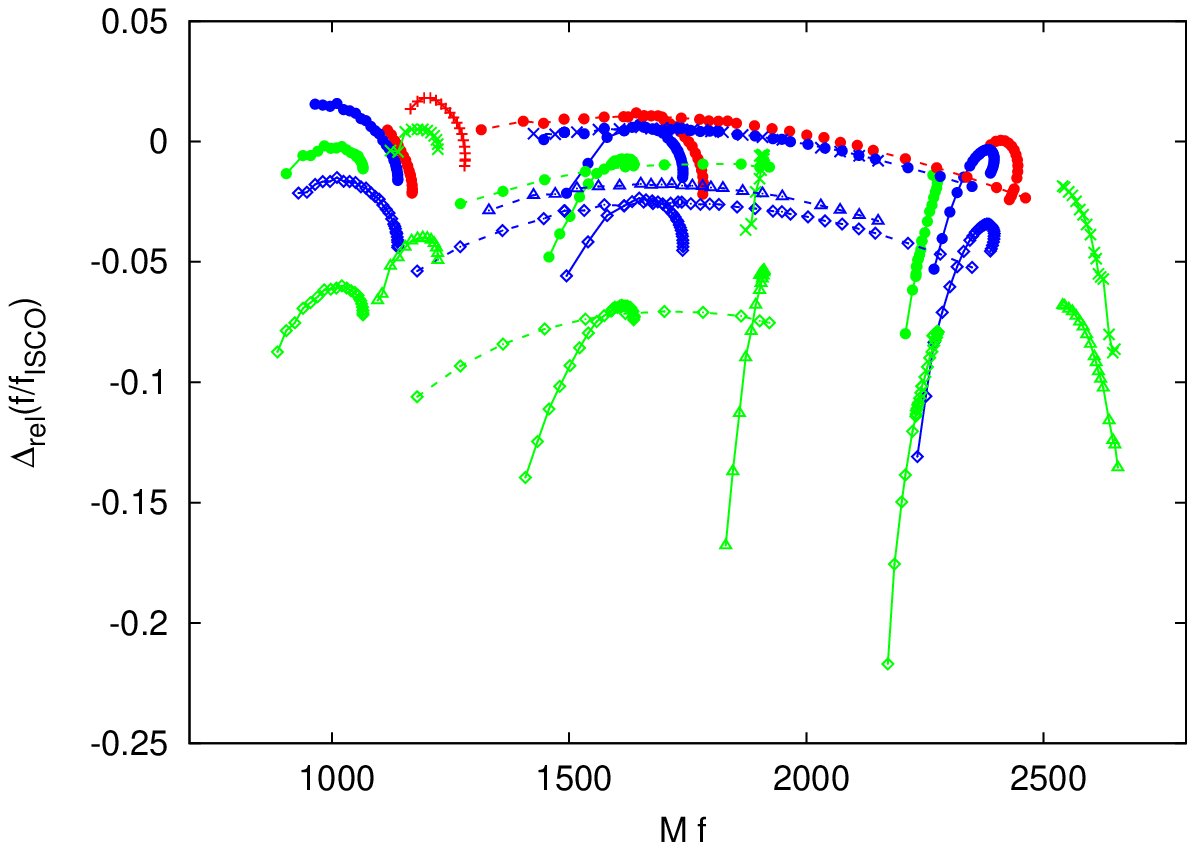}
}
}
\end{center}
\vspace{-0.5cm}
\caption{
Scaled ISCO radius $R_{\rm ISCO} f$ (upper left)
and scaled ISCO frequency $f/f_{\rm ISCO}$ (lower left)
versus the scaled neutron star mass $M f$,
for GR (red), dEGB with $\alpha=1.0$ (blue) and
dEGB with $\alpha=2.0$ (green),
with EOS FPS,
matter coupling $\beta=0$ (dots) and $\beta=0.5$ (diamonds),
as well as EOS DI-II, for $\beta=0$ (crosses), and $\beta=0.5$ (triangles).
Also shown are the relative deviations from the GR fits 
for the scaled ISCO radius (upper right)
and the scaled ISCO frequency (lower right).
}
\label{Fig8}
\end{figure}

\begin{figure}[h!]
\begin{center}
\mbox{\hspace{0.2cm}
{\hspace{-1.0cm} 
\includegraphics[height=.25\textheight, angle =0]{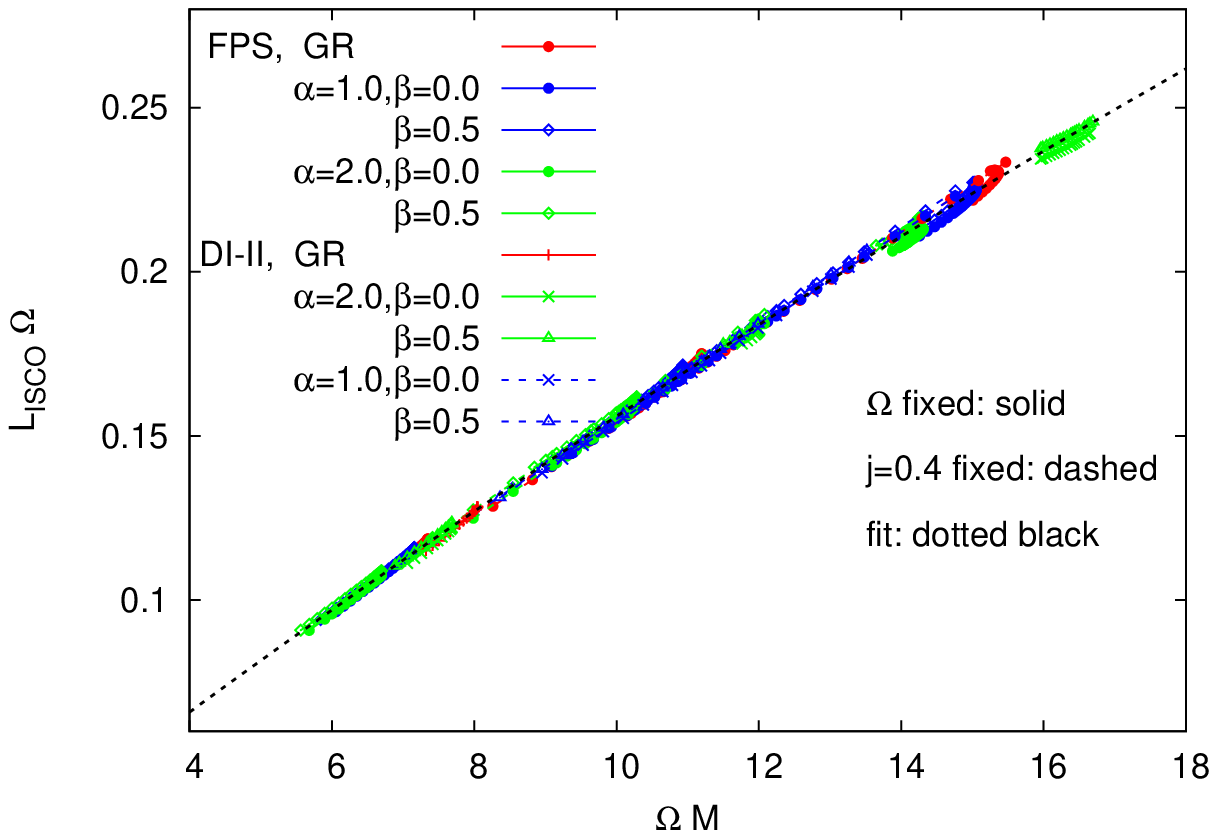}
}
{\hspace{-0.5cm}
\includegraphics[height=.25\textheight, angle =0]{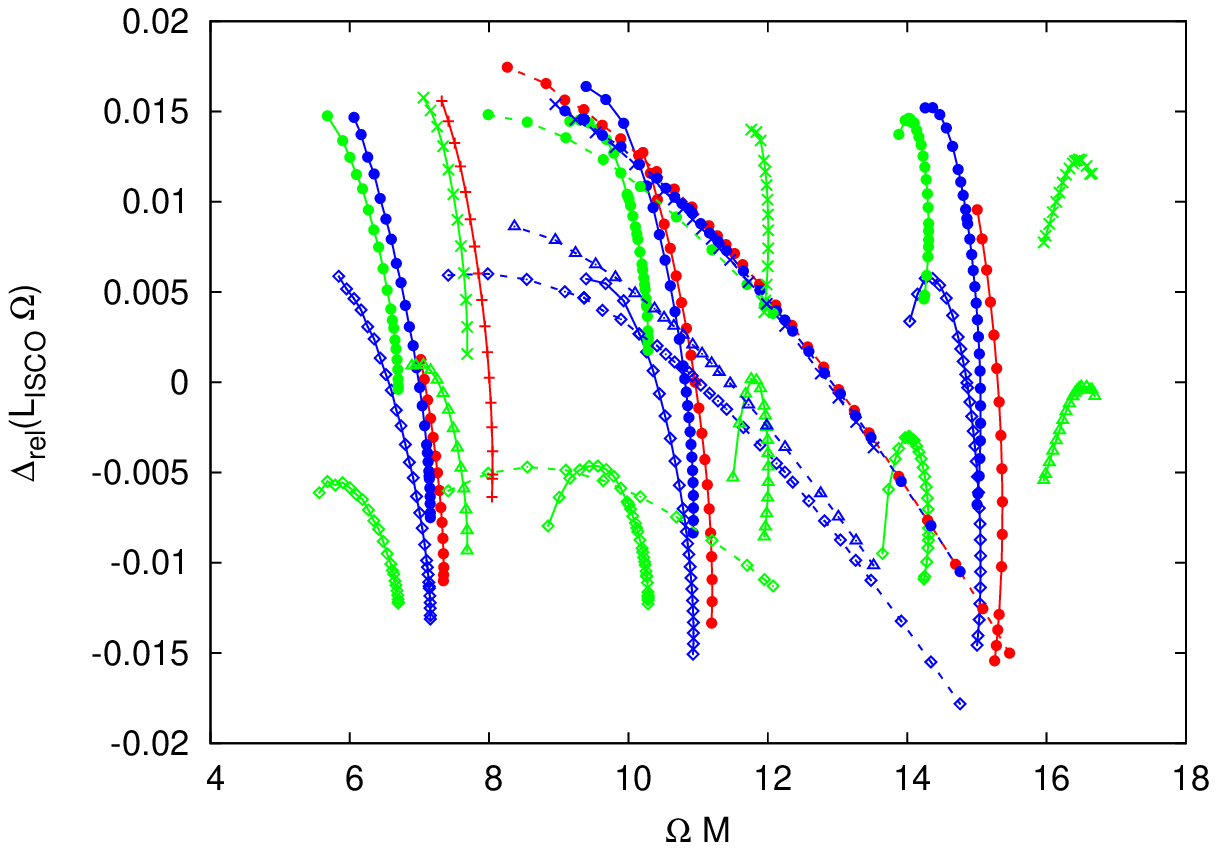}
}
}
\end{center}
\vspace{-0.5cm}
\caption{Scaled ISCO angular momentum $L_{\rm ISCO} \Omega$
versus the scaled neutron star mass $M \Omega$ (left)
and deviations from the fit (right).
The data consist of both EOS, GR and dEGB with $\alpha=1$, 2,
$\beta=0$, $0.5$.
}
\label{Fig9}
\end{figure}

We illustrate the full set of data and their deviations from 
their respective fits in Fig.~\ref{Fig9}. 
Together with the data points shown in Fig.~\ref{Fig9} for the 
scaled ISCO radius $R_{\rm ISCO} \Omega$ (upper left)
and the scaled ISCO frequency $f/f_{\rm ISCO}$ (lower left)
we exhibit the respective curves for the best fit for GR \cite{Luk:2018xmt},
for dEGB with $\alpha=1$, $\beta=0.5$, and
for dEGB with $\alpha=2$, $\beta=0$. 
The deviations from the corresponding best fits 
for the scaled ISCO radius $R_{\rm ISCO} \Omega$ (upper left)
and the scaled ISCO frequency $f/f_{\rm ISCO}$ (lower left)
are also exhibited in the figure.
While the universality remains good for dEGB with $\alpha=1$
and $\beta=0$ (deviations of 4\%), 
it clearly deteriorates for dEGB with $\alpha=1$, $\beta=0.5$ 
(deviations of 8\%), $\alpha=2$, $\beta=0$ (deviations of 8\%),
and becomes broken for dEGB with $\alpha=2$, $\beta=0.5$,
where deviations reach 25\%.

Thus the full universality seen in these relations for GR
and surviving to dEGB with $\alpha=1$ and $\beta=0$ 
is giving way to a restricted universality, where an additional
dependence on another dimensionless quantity, such as $j$,
is needed to describe the data properly, i.e., one should instead
consider a surface for extracting universal relations.
For such an analysis our data are, however, insuffcient. 
We note, that a large part of our data are based on constant values 
of the dimensionful quantity $\Omega$.
We also note, that such a structure would then be rather similar 
to the structure of the $I$-Love-$Q$ relations studied previously
\cite{Doneva:2013rha,Pappas:2013naa,Chakrabarti:2013tca}.

Considering the above results for the universal relations for 
$R_{\rm ISCO}f$ and $f/f_{\rm ISCO}$,
it is all the more surprising that there exists a full universal relation
for another ISCO property, namely for the scaled ISCO angular momentum
$L_{\rm ISCO} \Omega$.
This new universal relation is shown in Fig.~\ref{Fig9}, 
where all of the properly scaled data is superimposed.
Clearly the data for both EOS, for GR, for dEGB with $\alpha=1$ and 2,
as well as $\beta=0$ and $0.5$ all can be fitted by a single
curve, and the deviations from this fit do not exceed 3\%.
Let us recall here for comparison Fig.~\ref{Fig6}, 
which shows the unscaled quantities with their various dependencies.
Thus the scaling removes all of these dependencies,
resulting in a single universal curve.
The universal relation reads
$$ L\Omega_{\rm ISCO} = c_1 M \Omega + c_2 (M \Omega)^2 + c_3 (M \Omega)^3$$
with 
$c_1=1.707 \times 10^{-2} \ , \
 c_2=-1.567 \times 10^{-4} \ , \
 c_3= 9.147\times 10^{-7}$.

\begin{figure}[h!]
\begin{center}
\mbox{\hspace{0.2cm}
{\hspace{-1.0cm} 
\includegraphics[height=.25\textheight, angle =0]{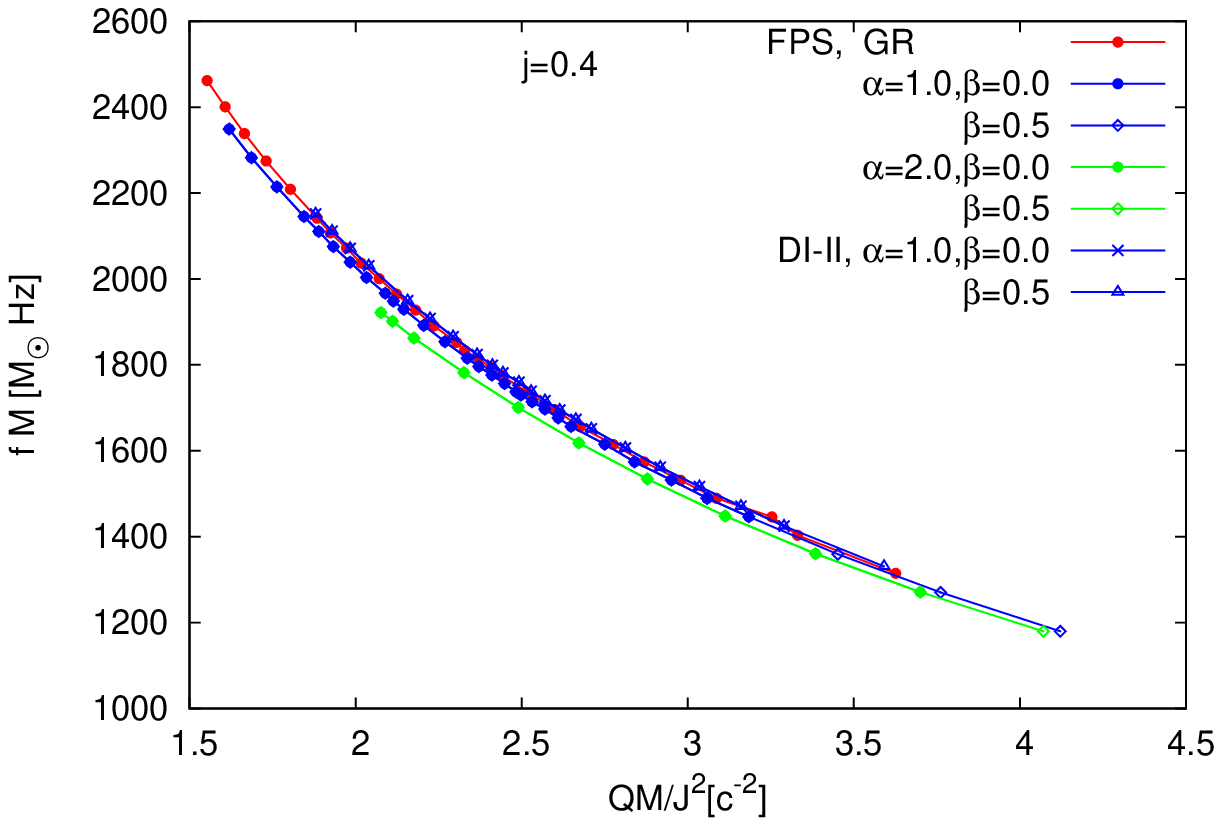}
}
{\hspace{-0.5cm}
\includegraphics[height=.25\textheight, angle =0]{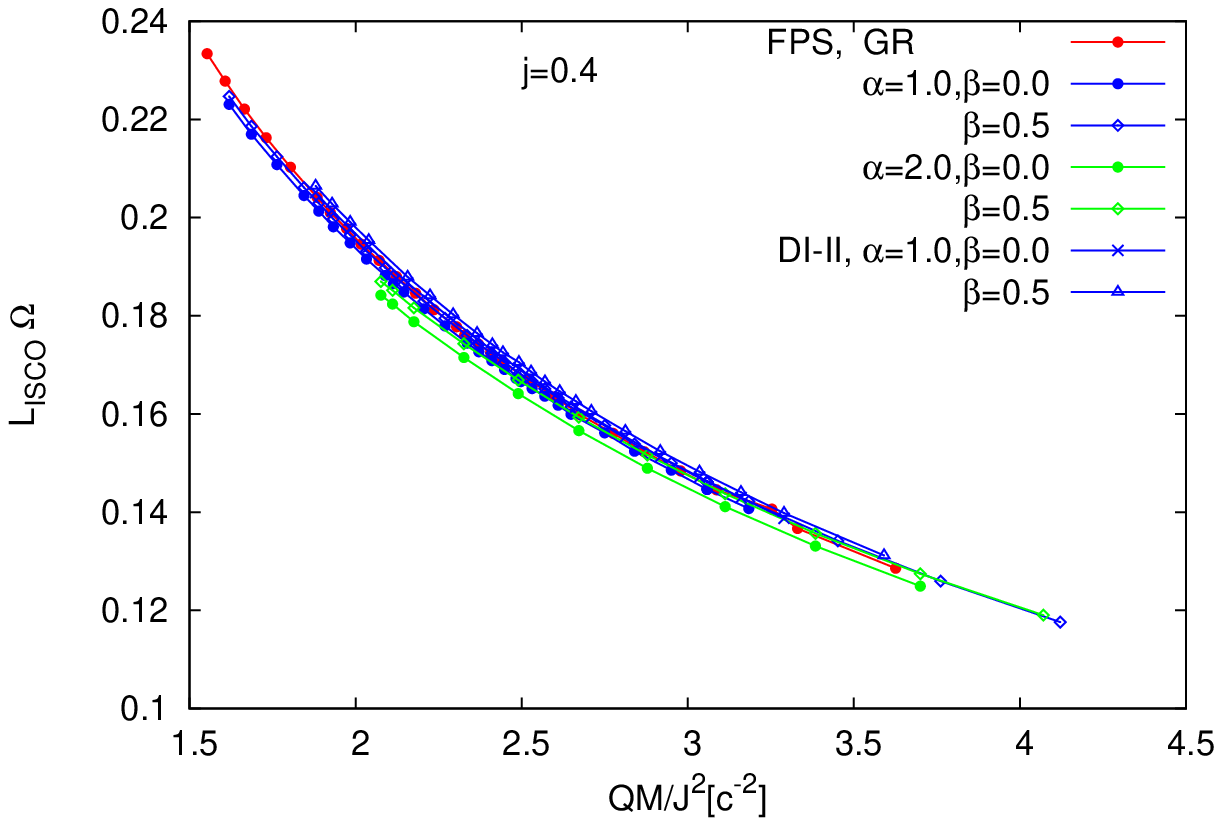}
}
}
\end{center}
\vspace{-0.5cm}
\caption{Scaled neutron star mass $M f$ (left)
and scaled ISCO angular momentum $L_{\rm ISCO} \Omega$ (right)
versus the scaled quadrupole moment $QM/J^2$
for a fixed value of the scaled angular momentum $j=0.4$ of the star,
matter coupling $\beta=0$ (dots) and $\beta=0.5$ (diamonds),
EOS FPS, for GR (red), dEGB with $\alpha=1.0$ (blue) and
dEGB with $\alpha=2.0$ (green),
as well as EOS DI-II, for dEGB with $\alpha=1.0$, $\beta=0$ (crosses),
and $\beta=0.5$ (triangles).
}
\label{Fig10}
\end{figure}

Let us finally connect to previous universal relations
\cite{Yagi:2016bkt,Doneva:2017jop},
and, in particular, to the universal $I$-$Q$ relation in dEGB theory 
\cite{Kleihaus:2014lba,Kleihaus:2016dui}.
This relation is known to depend on the dimensionless angular
momentum $j$, and thus corresponds to a universal surface.
Let us now consider a particular curve on this surface given by a
fixed value of $j$. Then there is almost no dependence on the EOS 
and the parameters $\alpha$ and $\beta$ 
\cite{Kleihaus:2014lba,Kleihaus:2016dui}.
If we consider instead the scaled mass $M \Omega$
versus the scaled quadrupole moment $QM/J^2$,
then this high degeneracy with respect to
$\alpha$ and $\beta$ is slightly broken, 
as seen in Fig.~\ref{Fig10} (left).

Not surprisingly, this slight breaking of the degeneracy
is also visible, when the scaled ISCO angular momentum $L \Omega$ is
considered versus the scaled quadrupole moment $QM/J^2$,
as shown in Fig.\ref{Fig10} (right).
We note, that for the dependence of the
scaled ISCO radius $R_{\rm ISCO} \Omega$ or the
scaled ISCO frequency $\Omega_c/\Omega$ 
on the scaled quadrupole moment $QM/J^2$,
the breaking of the degeneracy with $\alpha$ and $\beta$
increases as compared to $L \Omega$.
In all cases, universal surfaces are expected to result, 
when data with different values of $j$ are included.

\begin{figure}[h!]
\begin{center}
\mbox{\hspace{0.2cm}
{\hspace{-1.0cm}
\includegraphics[height=.25\textheight, angle =0]{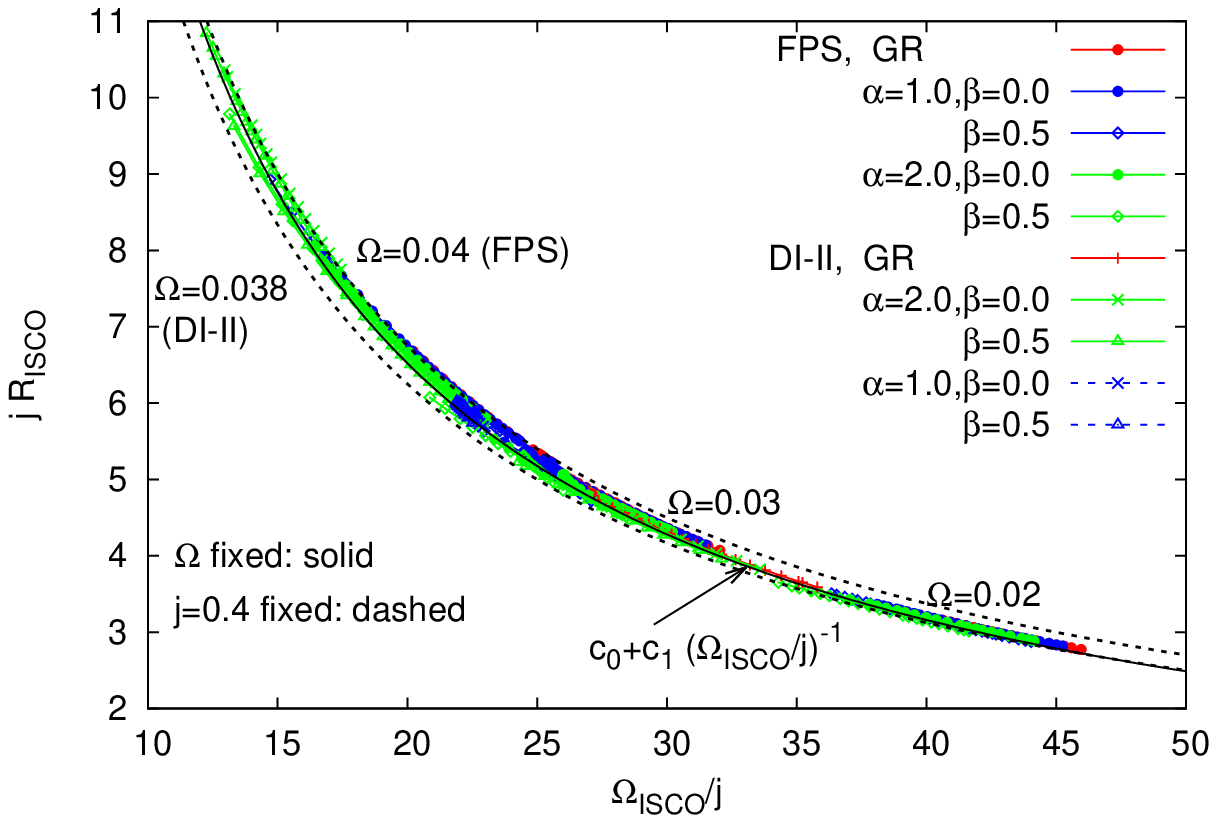}
}
{\hspace{-0.5cm}
\includegraphics[height=.25\textheight, angle =0]{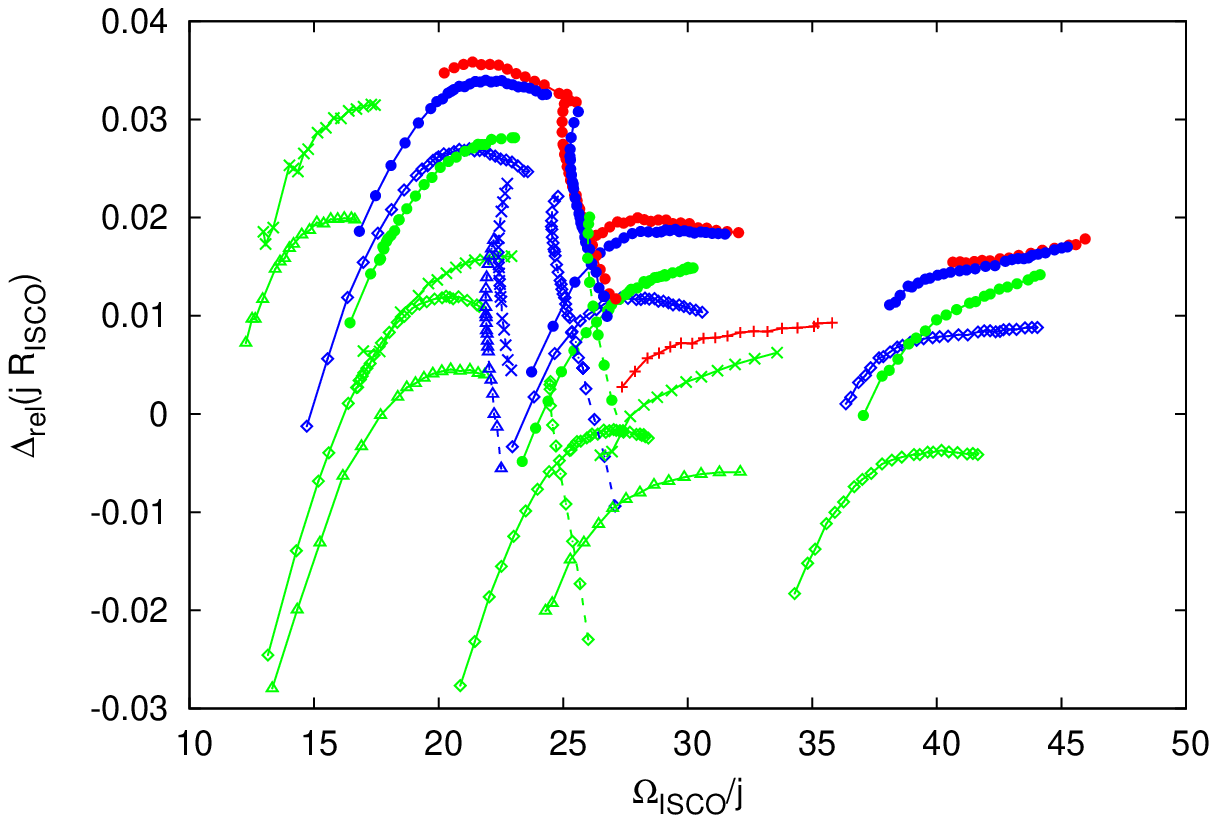}
}
}
\end{center}
\vspace{-0.5cm}
\caption{Scaled ISCO radius $R_{\rm ISCO} j$ 
versus the scaled ISCO angular velocity $\Omega_{\rm ISCO}/j$ (left)
and deviations from the fitted curve (right).
The data consist of both EOS, GB and dEGB with $\alpha=1$, 2,
$\beta=0$, $0.5$.
}
\label{Fig11}
\end{figure}

As a last point of interest let us mention, that we have also
observed a relation, that looks like a full universal relation,
since a single fit covers all the data within 3\% accuracy,
thus including the variations of $\alpha$ and $\beta$,
$$ R_{\rm ISCO} j = c_0 + c_1 \left(\Omega_{\rm ISCO}/j\right)^{-1} \ , $$
with $c_0=-0.2$ and $c_1=134.4$.
However, this relation involves dimensionful quantities,
the ISCO radius $ R_{\rm ISCO}$ and ISCO angular velocity $\Omega_{\rm c}$, 
scaled by the dimensionless angular momentum $j$.

\section{Conclusions}

Since neutron stars are highly compact objects,
they represent - in principle - a means 
to extract a lot of information about GR 
or generalized theories of gravity. 
However, there is a price to pay, 
since the properties of neutron stars
are highly dependent on the EOS.
Here universal relations represent an excellent means
to eliminate this EOS dependence to a large
extent. 

While universal relations for properties of
the neutron stars themselves have been studied
in much detail before, a set of universal relations
for the ISCOs of neutron stars in GR has 
only recently been announced \cite{Luk:2018xmt}.
These universal relations concern the
dependence of the scaled ISCO radius $R_{\rm ISCO} \Omega$
and the scaled ISCO frequency $\Omega_{\rm c}/\Omega$
on the scaled neutron star mass $M \Omega$.
Here we have added a further such relation,
namely the dependence of the scaled ISCO angular momentum
$L_{\rm ISCO}\Omega$ on the scaled mass $M \Omega$.

However, our main goal has been to gain some
understanding on how these universal relations are
affected, when a generalized theory of gravity
is employed instead of GR. Therefore we have
analyzed our neutron star data obtained in
the string theory motivated dEGB gravity, 
with GB coupling constants $\alpha=1$ and 2, 
allowing at the same time
for varying values of the matter coupling constant
$\beta=0$ and $\beta=0.5$, with the latter value
corresponding to the value suggested by string theory.

Inspection of the GR universal relations 
$R_{\rm ISCO} \Omega$ versus $M \Omega$ 
and $\Omega/\Omega_{\rm ISCO}$ versus $M \Omega$
for dEGB has shown, that a small dependence on
the GB coupling constant arises, as well as a
larger dependence on the matter coupling constant $\beta$.
Moreover, we conclude, that the single degenerate
curve should in principle be replaced by a surface,
where the scaled angular momentum $j$ could represent
the additional axis.

Interestingly, the new GR universal relation $L_{\rm ISCO}\Omega$ versus
$M \Omega$ remains basically untouched, as
the GB coupling constant $\alpha$ and the matter 
coupling constant $\beta$ are varied.
Indeed, for all available data the relative deviation 
from the GR fit does not exceed 3\%.
Consequently this new universal relation does not
discriminate between GR and dEGB theory, whereas the two
previously found relations do, when $\alpha$ or $\beta$
is sufficiently large.

There has also been previous work concerning relations
between ISCO properties and the multipole moments of 
neutron stars \cite{Shibata:1998xw,Pappas:2015mba,Luk:2018xmt}.
In particular, expansions have been considered,
relating the ISCO radius and the ISCO frequency
to dimensionless parameters like the scaled angular momentum $j$, 
the scaled quadrupole moment and the scaled spin octupole moment.
Here we have considered the dependence of $L_{\rm ISCO} \Omega$,
$R_{\rm ISCO} \Omega$ and $\Omega/\Omega_{\rm ISCO}$
on the scaled quadrupole moment $QM/J^2$.
In all cases, we observe for fixed $j$
a small dependence on the dEGB parameters,
and we expect universal surfaces when $j$ is varied as well.

As a final curious finding we note, that we have also encountered
a relation for dimensionful quantities, that otherwise looks
like a universal relation, valid both for GR and dEGB.
Here we have considered $R_{\rm ISCO} j$ versus $j/\Omega_{\rm ISCO}$.
Thus the ISCO properties are dimensionful, while they are scaled
with the dimensionless $j$. The deviations from the fitted curve
are below 4\% for this intriguing relation.

If observations would indicate sufficiently strong deviations 
from the GR universal relations, this would suggest the need 
for a more general theory of gravity, allowing for such deviations.
We have seen, that dEGB theory can accomplish deviations for certain GR
universal relations, while remaining in perfect agreement with 
certain other GR universal relations.
Viewing dEGB theory as a particular case of Horndeski gravity 
\cite{Kobayashi:2011nu} then suggests
to extend the current set of investigations
in order to address further cases of Horndeski gravity in the future.

Here one might, in particular, focus on candidate theories
that are also in good agreement with gravitational waves tests at the 
cosmological level \cite{Ishak:2018his}.
We note, that there are already various interesting results
on geodesic motion and accretion processes of black holes 
in Horndeski theories \cite{Tretyakova:2017rlk,Salahshoor:2018plr},
while neutron stars have also been investigated
in Horndeski theories with a focus on their global properties, their modes
and the associated universal relations (see e.g.
\cite{Cisterna:2015yla,Maselli:2016gxk,Blazquez-Salcedo:2018tyn,
Blazquez-Salcedo:2018pxo}). 

Let us end with some comments concerning proposed applications.
The ISCO may be related to QPOs of low mass x-ray binaries, 
where the QPOs are attributed to oscillations close to the
edge of the accretion disk, that is given by the respective ISCO
(see e.g.,~\cite{vanderKlis:2005,vanDoesburgh:2018oom}).
When the frequency of the neutron star and QPO frequency
associated with the ISCO are known, then the GR universal relation
would allow a determination of the neutron star mass
\cite{Luk:2018xmt}. For dEGB gravity on the other hand, this
simple relation would be lost, since with $\alpha$ and $\beta$
two unknown parameters enter the universal relations.
If, however, one could find a way to observationally extract the ISCO
angular momentum, then the neutron star mass could also
be determined in dEGB gravity.

\section*{Acknowledgements}

This research was partially supported by VIE-UIS, under
grant number 2416, and COLCIENCIAS grant number
110277657744 (VIE-UIS 8863), Colombia. SM wants to
thank the support from the Postdoctoral Fellowship Scheme
provided by VIE-UIS.
BK and JK gratefully acknowledge support by the
DFG Research Training Group 1620 {\sl Models of Gravity}
and the COST Action CA16104.

\end{document}